\documentclass[11pt]{article}

\usepackage{epsf,latexsym,amssymb, graphicx}

\newcommand{\journal}[1]{}
\newcommand{\preprint}[1]{#1}

\journal{}

\newcommand{\s}{{\bf s}}
\newcommand{\rt}{{\bf v}}
\newcommand{\pr}{{\bf p}}
\newcommand{\N}{{\bf N}}
\newcommand{\NR}{{\bf N_R}}
\newcommand{\Li}{{\bf L}}
\newcommand{\Lio}{{\bf L_0}}
\newcommand{\T}{{\bf T}}
\newcommand{\Rs}{{\bf R^s}}
\newcommand{\Rv}{{\bf R^J}}
\newcommand{\Cs}{{\bf C^s}}
\newcommand{\Cv}{{\bf C^J}}
\newcommand{\Ds}{{\bf D^s}}
\newcommand{\Dv}{{\bf D^v}}
\newcommand{\Dp}{{\bf D^p}}

\newcommand{\comment}[1]{}
\newcommand{\into}{\rightarrow}

\newcommand{\R}{{\mathbb R}}  
\newcommand{\rref}[1]{(\ref{#1})}

\newtheorem{theorem}{Theorem}
\newtheorem{itlemma}{Lemma}[section] 
\newtheorem{itproposition}[itlemma]{Proposition}
\newtheorem{itcorollary}[itlemma]{Corollary}
\newtheorem{itremark}[itlemma]{Remark}
\newtheorem{itdefinition}[itlemma]{Definition}
\newtheorem{itexample}[itlemma]{Example}
\newenvironment{lemma}{\begin{itlemma}\rm}{\end{itlemma}} 
\newenvironment{remark}{\begin{itremark}\rm}{\end{itremark}} 

\newenvironment{proposition}{\begin{itproposition}\rm}{\end{itproposition}}
\newenvironment{definition}{\begin{itdefinition}\rm}{\end{itdefinition}}
\newenvironment{example}{\begin{itexample}\rm}{\end{itexample}}
\def\bi{\begin{itemize}}
\def\ei{\end{itemize}}
\def\ben{\begin{enumerate}}
\def\een{\end{enumerate}}
\def \beq {\begin{eqnarray}}
\def \eeq {\end{eqnarray}}
\def \beqn {\begin{eqnarray*}}
\def \eeqn {\end{eqnarray*}}

\newcommand{\text}[1]{\hbox{\rm \ #1\ \/}}
\newcommand{\be}[1]{\begin{equation}\label{#1}}
\newcommand{\ee}{\end{equation}}
\newcommand{\bl}[1]{\begin{lemma}\label{#1}}
\newcommand{\br}[1]{\begin{remark}\label{#1}}
\newcommand{\bt}[1]{\begin{theorem}\label{#1}}
\newcommand{\bd}[1]{\begin{definition}\label{#1}}
\newcommand{\bp}[1]{\begin{proposition}\label{#1}}
\newcommand{\bc}[1]{\begin{itcorollary}\label{#1}}
\newcommand{\ec}{\mybox\end{itcorollary}}
\newcommand{\ecs}{\end{itcorollary}}
\newcommand{\bfact}[1]{\begin{fact}\label{#1}}
\newcommand{\bex}[1]{\begin{example}\label{#1}}
\newcommand{\bem}[1]{\begin{example}\label{#1}}  
\newcommand{\efact}{\mybox\end{fact}}
\newcommand{\eex}{\mybox\end{example}}
\newcommand{\eem}{\mybox\end{example}}
\newcommand{\el}{\mybox\end{lemma}}
\newcommand{\ele}{\mybox\end{lemmaex}}
\newcommand{\er}{\mybox\end{remark}}
\newcommand{\et}{\qed\end{theorem}}
\newcommand{\ed}{\mybox\end{definition}}
\newcommand{\ep}{\mybox\end{proposition}}
\newcommand{\epr}{\end{proof}}
\newcommand{\bpr}{\begin{proof}}

\newcommand{\eers}{\end{exercise}}
\newcommand{\eexs}{\end{example}}
\newcommand{\eems}{\end{example}}
\newcommand{\els}{\end{lemma}}
\newcommand{\eles}{\end{lemmaex}}
\newcommand{\ers}{\end{remark}}
\newcommand{\ets}{\end{theorem}}
\newcommand{\eds}{\end{definition}}
\newcommand{\eps}{\end{proposition}}
\newcommand{\qed}{\hfill \halmos} 
\newcommand{\mybox}{\hfill $\Box$} 
\newcommand{\halmos}{\rule{1ex}{1.4ex}}
\newenvironment{proof}{\noindent {\em Proof}.\ }{\hspace*{\fill}$\halmos$\medskip}

\newcommand{\ve}{\varepsilon }

\def\edo{\end{document}}
\renewcommand{\tilde}{\widetilde}
\renewcommand{\hat}{\widehat}

\comment{possible referees:
Hofmeyr, W'Hoff, Boris, Fell
}

\begin{document}

\journal{\noindent {\bf Sensitivity Analysis of Stoichiometric Networks: An Extension of
Metabolic Control Analysis to Non-equilibrium Trajectories}
\\
\\

\noindent Running Title: MCA of Non-equilibrium Trajectories
\\
\\

\noindent Summary: 
A sensitivity analysis of general stoichiometric networks is considered.  
The 
results are presented as a generalization of Metabolic Control Analysis, which
has been concerned primarily with system sensitivities at steady state.
An expression for time-varying sensitivity coefficients is given, and
the Summation and Connectivity Theorems
are generalized.  The results are compared to previous treatments.  
The analysis is accompanied by a 
discussion of the computation
of the sensitivity coefficients and an application to a model of 
phototransduction.}

\title{Sensitivity Analysis of Stoichiometric Networks: An Extension of
Metabolic Control Analysis to Non-equilibrium Trajectories}
\author{Brian P. Ingalls\thanks{Supported by Air Force Research 
Laboratory Cooperative Agreement No. F30602-01-2-0558} \
and Herbert M. Sauro\thanks{Supported by Air Force Research 
Laboratory Cooperative Agreement No. F30602-01-2-0558 and Japan Sciences and
Technology Corporation (JST) Kitano ERATO Project} \\ 
Control and Dynamical Systems, California Institute of Technology, CA
\\
{\tt \{ingalls,hsauro\}@cds.caltech.edu} 
}

\date{}
\maketitle

\begin{abstract}
\noindent
 A sensitivity analysis of general stoichiometric networks is considered.  
The 
results are presented as a generalization of Metabolic Control Analysis, which
has been concerned primarily with system sensitivities at steady state.
An expression for time-varying sensitivity coefficients is given, and
the Summation and Connectivity Theorems
are generalized.  The results are compared to previous treatments.  
The analysis is accompanied by a 
discussion of the computation
of the sensitivity coefficients and an application to a model of 
phototransduction.
\end{abstract}

\section{Introduction}

Sensitivity analysis is an important tool in the study  of
systems dependent on external parameters.  By their nature, stoichiometric 
networks allow an elegant description of the relationships 
between component and systemic sensitivities.  These relationships,
particularly those existing at equilibria of the system, have been 
addressed by the fields of Biochemical Systems Theory (BST)~\cite{savageau} 
and Metabolic Control
Analysis (MCA)~\cite{fell, heinrich-schuster}.

This paper describes the computation and interpretation of system
sensitivities over arbitrary trajectories.  The results are presented in
the framework provided by MCA, which is a theory devoted to 
the analysis of
the distribution of control within a network and how the behaviour of the
system relates to the properties of the components.  
With few exceptions
the MCA literature treats 
networks in steady state.

MCA has had great success in describing the control and regulation of
systems at steady state.  However, in the analysis of an increasing domain of examples
it becomes necessary to consider sensitivities along non-steady
trajectories.   In many systems it is the transient or oscillatory behaviour
which is of primary interest (e.g.~in signal transduction or
cell cycle regulation). 
Some extensions of MCA to non-steady
behaviour have appeared in the literature
(\cite{ask, demin, reder-heinrich, Kho-periodic, kohn, reijenga}).  
These papers generalize the steady state results of MCA to special cases of
dynamic behaviour (e.g.~periodic behaviour or 
trajectories near a stable equilibrium).
The analysis presented in this paper 
extends these results by measuring time varying sensitivities
along arbitrary trajectories.

A standard MCA analysis might treat, for example,
the effect of increasing an enzyme concentration on the steady state value
of some related metabolite.  
Using the analysis described below, one can determine 
the sensitivity to the perturbation {\em throughout the time evolution} of the 
system, regardless of the nature of the trajectory.  Based on 
these time-varying sensitivities, the
Summation and Connectivity Theorems of MCA are extended to give conditions
which hold for all time.
When applied at steady state, these results reduce to the
standard analysis of MCA.  

The statements in this paper 
apply to arbitrary system dynamics, including transients, 
convergence to steady state, and oscillations.  However, as pointed out
in previous papers (\cite{demin, Kho-periodic}), 
sensitivity coefficients derived for oscillating systems 
must be interpreted with care.  A discussion on applications to 
oscillating systems appears below.

The outline of the paper is as follows.
  Time-varying sensitivity coefficients are 
defined and derived, and their computation is discussed briefly.
The main results of MCA -- the Summation and Connectivity Theorems -- are then
extended to the time-varying case.
Finally, the sensitivity analysis is illustrated by treating some examples, 
including a model of phototransduction.

\section{Preliminaries}

A network consisting of $n$ chemical species involved in $m$ 
reactions is modelled.  The $n$-vector $\s$ is composed of the concentrations
of each species.  The constant $r$-vector $\pr$ is composed of the (external)
 parameters
of interest in the model.  The $m$-vector valued 
function $\rt = \rt(\s, \pr, t)$
describes the rate of each reaction as a (possibly time-varying) 
function of species 
concentrations and parameter values.  Finally the $n$ by $m$ 
stoichiometry matrix $\N$ 
describes the network: component $\N_{i,j}$ is equal to
the net number of individuals of species $i$ produced or consumed 
in reaction $j$.  The 
network can then be modelled by the ordinary differential equation
\beq
\label{sys}
\frac{d}{dt} \s(t) = \N \rt(\s(t),\pr, t) \qquad \text{for all } t \geq 0. 
\eeq
We allow the vector $\pr$ to
contain the initial states of any of the species
concentrations as well as any other parameters which will directly
affect the rates of the reactions (e.g.~concentrations of enzymes and external
effectors).

We assume that the function $\rt(\s, \pr, t)$ is continuous and is 
continuously 
differentiable in $\s$ and $\pr$ for each fixed $t$.  We further assume
 that for each initial condition $\s(0)=\s_0$ and each choice of 
parameters $\pr$ 
the unique solution
of~\rref{sys} is defined for all $t \geq 0$.  
We denote this solution 
by $\s(t, \s_0, \pr)$, and will use simply $\s(t, \pr)$ or 
$\s(t)$ when no confusion
will arise.

Before embarking on an analysis of system~\rref{sys} it is prudent to first
consider any linear dependencies inherent in the state variables of 
the system (which will simplify both the analysis and the computation).  
Each conserved moiety in the
network corresponds to a linearly dependent row in the stoichiometry matrix 
$\N$.  We follow the procedure and terminology described by 
Reder~\cite{reder88} (see also~\cite{hofmeyr-nutshell}) in the reduction
of the system.

Let $n_0$ denote the row rank of $\N$.  If $n_0=n$, then no reduction is
necessary.  Otherwise, we 
begin by re-ordering the rows of $\N$ so that the first $n_0$ rows are 
linearly independent.  Let $\NR$ be the reduced stoichiometry matrix which
results from truncating the last $n-n_0$ rows of $\N$.  
Since the truncated rows
can be formed by linear combination of the rows of $\NR$, the matrix $\N$ can
be written as the product
\beqn
\N = \Li \NR
\eeqn
where the $n \times n_0$ matrix $\Li$, called the 
{\em link matrix}, has the form
\beqn
\Li= \left [ \begin{array}{c} {\bf I}_{n_0} \\ {\bf L_0} \end{array} \right ].
\eeqn
(Here and below the notation ${\bf I}_q$ 
will be used for the $q \times q$ identity
matrix).

The advantage of this decomposition can now be realized.  Since each conserved
moiety allows one species concentration to be determined as a function
of the others, we may decompose the species vector $\s$ into independent
and dependent species vectors $\s_i$ and $\s_d$ respectively.  
Ordering the components of $\s$ to match the rows of $\N$, we write
$\s = (\s_i, \s_d)$ where $\s_i$ is an $n_0$-tuple and $\s_d$ 
is an $(n-n_0)$-tuple. (This involves a minor abuse of notation since
$\s$, $\s_i$ and $\s_d$ are all column vectors.)
Then equation~\rref{sys} can be written as
\beqn
\frac{d}{dt} \left [ \begin{array}{c} \s_i(t) \\ \s_d(t) \end{array} \right ]
= \Li \NR \rt(\s(t), \pr, t) 
= \left [ \begin{array}{c} {\bf I}_{n_0} \\ {\bf L_0} \end{array} \right ]
\NR \rt(\s(t), \pr, t) \qquad \text{for all } t \geq 0.
\eeqn
Hence
\beqn
\frac{d}{dt} \s_d(t) = {\bf L_0} \frac{d}{dt} \s_i(t)
 \qquad \text{for all } t \geq 0,
\eeqn
and so $\s_d(t) - {\bf L_0} \s_i(t)$ 
is an integral of motion of~\rref{sys}, i.e.~this 
difference is constant throughout the evolution of the system.  We introduce
the constant $(n-n_0)$-vector $\T$ to quantify this relationship.
Any trajectory of the system satisfies
\beq
\label{spec-dep}
\s_d(t) = \Lio \s_i(t) + \T \qquad \text{for all } t \geq 0,
\eeq
where $\T$ is defined in terms of the initial conditions as
\beqn
\T = \s_d(0) - \Lio \s_i(0).
\eeqn

In analysis and computation,  attention can be restricted 
to the independent species
$\s_i$, since the corresponding results incorporating the dependent vector
$\s_d$ are arrived at immediately through the relationship~\rref{spec-dep}.
That is, one need only consider the reduced system
\beq
\frac{d}{dt} \s_i(t) = \NR \rt(\s(t),\pr,t) = 
\NR \rt((\s_i(t), \Lio \s_i(t) + \T), \pr, t) \qquad \text{for all } t \geq 0.
\label{redsys}
\eeq

\section{Sensitivity Analysis}

We now present a general sensitivity analysis of the
system~\rref{sys}.

\subsection{Definitions}

Following~\cite{tomovic} and~\cite{ask}, we make the basic definition.

\bd{conc-response}  Given an initial condition $\s(0)= \s_0$ and 
a set of parameter values
$\pr_0$, we define the time-varying 
{\em concentration sensitivity coefficients} (or
{\em concentration response coefficients}) as the elements of the
$n \times r$ matrix function
$\Rs(\cdot)$ given by
\beqn
\Rs(t) := \frac{\partial \s(t, \pr)}{\partial \pr} |_{\pr=\pr_0}
= \lim_{\Delta \pr \into 0}
\frac{\s(t, \pr_0 + \Delta \pr) - \s(t, \pr_0)}{ \Delta \pr} \qquad 
\text{for all } t \geq 0.
\eeqn
\ed

\comment{
The sensitivity coefficients are well defined for all t and all $\pr_0$ 
since the function $\rt(\s, \pr, t)$ 
is assumed continuously differentiable in $\pr$.}

\br{Rs-opmeaning}
These time varying response coefficients can be interpreted in exact analogy
to the standard (steady state) response coefficients of MCA.  The 
difference is that the response to a perturbation along an entire trajectory
in now considered, rather than at a particular equilibrium state.  

Take for example a system which has an asymptotically stable equilibrium 
(say $\s^{ss}$), and
consider a set of initial conditions (say $\s_0$) 
which do not match the steady state
solution.  Letting the system evolve
from $\s_0$, we may observe some initial
transient, followed by convergence to $\s^{ss}$ as time tends to infinity.  
Alternatively, if we make a small perturbation to a system parameter,
we may observe a different transient, 
followed by convergence to a different steady state.  The response
coefficient defined above provides a measure of the difference between this
``perturbed trajectory'' and the ``nominal'' (unperturbed) trajectory at each 
time $t$.  As time tends to infinity, each trajectory will converge to its 
steady state, and so the response coefficient will converge to the steady
state response of MCA.  

This time-history analysis is particularly useful when studying systems in
which transient behaviour plays a key role in the mechanism.  For example,
 in systems 
performing phototransduction or action potential transfer, the role of
the system is to produce a (transient) spike in the concentration of a certain
species.  The analysis presented here will allow elucidation of the effect of
parameter perturbations on the behaviour of such a system.

It should be noted that the response coefficient $\Rs(\cdot)$ is defined 
with respect to a particular parameter choice $\pr_0$ and a particular 
initial condition $\s_0$.  Should one be interested in comparing the system
response to {\em perturbations at different times}, a separate response
coefficient must be defined for each such time, since each choice will
yield a distinct ``initial'' state (by re-setting time to zero
at the point chosen).
\er

\br{moiety totals}
Since the vector $\pr$ is allowed to contain the initial states as components,
sensitivity coefficients with respect to initial conditions are included in
the above definition.  
When pursuing a steady state analysis, it is often more convenient to consider
the total amount of each conserved moiety as a parameter rather than the initial
conditions of particular species 
(see e.g.~\cite{hofmeyr-nutshell}).  However, in this
dynamic analysis, perturbation of the initial concentrations of
different species will have distinct effects on the time-history; 
it is only at steady state
that these effects can be described equivalently 
in terms of perturbations in pools of conserved moieties.
\er

In addition to the sensitivities of the species concentrations,
it is also of interest to consider the sensitivities of the various rates 
$\rt(\s,\pr,t)$.  In expressing these responses, we conform to the 
MCA usage of the term {\em flux} for the rate of a system reaction.  
(In general, these
rates do not represent {\em fluxes} in the usual sense of the word as  a
rate of transfer of some quantity; the rate of passage of mass through a 
reaction pathway must also take the stoichiometry into consideration.)

\bd{rate-response}  Given an initial condition $\s(0)=\s_0$ and
 set of parameter values
$\pr_0$, we define the time-varying 
{\em flux sensitivity coefficients} (or
{\em flux response coefficients}) as the elements of the
$m \times r$ matrix function
$\Rv(\cdot)$ given by
\beqn
\Rv(t) := \frac{\partial \rt(\s(t, \pr), \pr, t)}{\partial \pr} |_{\pr=\pr_0} 
& = &
\frac{\partial \rt(\s, \pr,t)}{\partial \s} 
 \frac{\partial \s(t, \pr)}{\partial \pr} 
+ \frac{\partial \rt(\s, \pr,t)}{\partial \pr}  \\
&=& \frac{\partial \rt(\s, \pr,t)}{\partial \s}
 \Rs(t)
+ \frac{\partial \rt(\s, \pr, t)}{\partial \pr}
\qquad \text{for all } t \geq 0,
\eeqn
where the derivatives are evaluated at $\pr=\pr_0$ and $\s=\s(t, \pr_0)$.
\ed

\br{Rv-op}
The flux response coefficients can be interpreted analogously to the 
concentration response coefficients: $\Rv(t)$ gives the response in the rates 
at time $t$ to a perturbation at time zero.  At steady state, these
coefficients reduce to their standard MCA counterparts.
\er

\subsection{Computation}

The sensitivity coefficients are defined by a first-order
linear ordinary differential
equation which
follows from taking the derivative of~\rref{sys} with respect
to $\pr$.  Dropping some 
functional dependencies for ease of legibility, we have: 
\beqn
\frac{\partial}{\partial \pr} \frac{d \s(t)}{dt} = 
\N \left ( \frac{\partial \rt(t)}{\partial \s}
\frac{\partial \s(t)}{\partial \pr} + 
\frac{\partial \rt(t)}{\partial \pr} \right) \qquad \text{for all } t \ge 0
\eeqn
which becomes, after switching the order of differentiation,
\beq
\label{full-senode}
\frac{d}{d t} \frac{\partial \s(t)}{\partial \pr}  = \N
\left ( \frac{\partial \rt(t)}{\partial \s} \frac{\partial \s(t)}{\partial \pr}
+ \frac{\partial \rt(t)}{\partial \pr} \right ) 
\qquad \text{for all } t \ge 0. 
\eeq
Recall that the sensitivity coefficients are defined with respect to a 
particular choice of $\s_0 = \s(0)$ and $\pr_0$.  This choice fixes a 
trajectory $\s(t)= \s(t, \s_0, \pr_0)$, and hence determines 
the function $\rt(t) = 
\rt(\s(t), \pr, t)$ as well.
The initial conditions for~\rref{full-senode} (i.e.~the components of
the matrix
$\frac{\partial \s}{\partial \pr}(0)$) 
are immediate from the definition of the sensitivity coefficients:  the
$j,k$-th entry of $\frac{\partial \s}{\partial \pr}(0)$ will be zero
unless the $k$-th parameter is the initial condition of the $j$-th
species, in which case the initial value of the sensitivity coefficient will
be one.
Equation~\rref{full-senode} can be solved for the concentration
sensitivities 
$\frac{\partial \s}{\partial \pr}(\cdot)
= \Rs(\cdot)$, and the flux sensitivities can be computed according to 
definition~\ref{rate-response}.  
However, we can reduce the burden of computation
by taking advantage of the dependencies described in equation~\rref{spec-dep}.

Differentiating~\rref{redsys} with respect to $\pr$, we find
\beq
 \frac{\partial}{\partial t} \frac{\partial
\s_i(t)}{\partial \pr}  &=&  
\NR \left( \frac{\partial \rt(t)}{\partial \s_i}
\frac{\partial \s_i(t)}{\partial \pr} + \frac{\partial \rt(t)}{\partial \s_d}
\frac{\partial}{\partial \pr} (\Lio \s_i(t) + \T) + 
\frac{\partial \rt(t)}{\partial \pr} \right ) 
\nonumber \\
 &=&  \NR  \left [ \left ( \frac{\partial \rt(t)}{\partial \s_i} + 
\frac{\partial \rt(t)}{\partial \s_d} \Lio \right )
\frac{\partial \s_i(t)}{\partial \pr}  +   \frac{\partial
\rt(t)}{\partial \s_d} \frac{\partial \T}{\partial \pr} + 
\frac{\partial \rt(t)}{\partial \pr}  \right ] \nonumber \\
 &=&  \NR  \left [  \frac{\partial \rt(t)}{\partial \s} \Li
\frac{\partial \s_i(t)}{\partial \pr}  +   \frac{\partial
\rt(t)}{\partial \s_d} \frac{\partial \T}{\partial \pr} + 
\frac{\partial \rt(t)}{\partial \pr}  \right ]
\qquad \text{for all } t \ge 0. 
\label{mainode}
\eeq
The initial conditions are determined as before.

\br{num}
While there are special cases in which the response coefficients can be 
derived explicitly (as discussed below), in most cases
the solution to equation~\rref{full-senode} 
(or more efficiently~\rref{mainode}) must be computed numerically.  
In performing such a calculation, one must keep in mind that the 
right-hand-side depends on the time-varying value of the 
species concentration vector $\s(t)$.  A simple computational strategy 
is to  solve~\rref{redsys}
and~\rref{mainode} as a single set of differential equations, 
determining $\s_i(\cdot)$ and 
$\frac{\partial \s_i(\cdot)}{\partial \pr}$ simultaneously.  This would
allow, for example, an adaptive integrator to reduce the step-size during 
computation should either equation demand it.  

The reader should note
that the pair of ODE's are coupled, but that the coupling is in cascade --
equation~\rref{redsys} is independent of 
$\frac{\partial \s_i(\cdot)}{\partial \pr}$.  Thus the increase
in complexity comes simply from having two equations to solve, not from
any intertwining of the solutions.

An algorithm for computation of the time-varying response coefficients has 
been implemented in MATLAB and in the biochemical simulator 
{\em Jarnac}~\cite{jarnac}.
Scripts of the implementation are available online (http://www.sys-bio.org).
\er

\subsection{Analytical Solution}
\label{analytic-sol}

The solution to equation~\rref{mainode} can be expressed in terms of the
variation of parameters formula (derived in standard texts on ODE's, 
e.g.~\cite{boyce-dip}):
\beq
\label{varpar}
\frac{\partial \s_i}{\partial \pr}(t) =  
\Phi(t)
\frac{\partial \s_i}{\partial \pr}(0) + 
\Phi(t) \int_0^t \Phi^{-1}(\tau) \NR \left ( \frac{\partial
\rt(\tau)}{\partial \s_d} \frac{\partial \T}{\partial \pr}
+ \frac{\partial \rt(\tau)}{\partial \pr} \right ) \, d\tau
\qquad \text{for all } t \ge 0,
 \eeq
where the matrix-valued function $\Phi(\cdot)$ (called the {\em
fundamental matrix} for equation~\rref{mainode}) satisfies the homogeneous
part of~\rref{mainode}, i.e. $\Phi(\cdot)$ is defined as the solution of the
following initial value problem
\beq
\label{homo}
\frac{d}{dt} \Phi(t) = \left [ \NR   \frac{\partial \rt(t)}{\partial \s} 
 \Li \right ] \Phi(t),
\qquad \Phi(0) = {\bf I}_{n_0}.
\eeq

Note that~\rref{varpar} does not in general provide an explicit
formula for the solution to~\rref{mainode} (since~\rref{homo} does not admit 
an explicit solution, in general).    Nevertheless, knowing the solution
takes this form can simplify both analysis and computation.  
In addition, as will be demonstrated in Section~\ref{thms}, 
there are special cases
in which~\rref{varpar} does indeed provide an explicit solution.

In the subsequent analysis, it will be convenient to partition the 
parameter vector $\pr$ into 
the set of parameters which 
affect the rates (denoted $\pr_v$) and the set 
of initial conditions (denoted $\pr_s$).
From the fact that $\frac{\partial \rt}{\partial \pr_s} = 
\frac{\partial \s_i}{\partial \pr_v}(0) = 
\frac{\partial \T}{\partial \pr_v} = 0$, we see that
formula~\rref{varpar} can be partitioned as
\beq
\label{varpars}
\nonumber 
\frac{\partial \s_i}{\partial \pr_s}(t) &=  &
\Phi(t)
\frac{\partial \s_i}{\partial \pr_s}(0) + 
\Phi(t) \int_0^t \Phi^{-1}(\tau) \NR  \frac{\partial
\rt(\tau)}{\partial \s_d} \frac{\partial \T}{\partial \pr_s} \, d\tau \\
\label{varparv}
\frac{\partial \s_i}{\partial \pr_v}(t) &= &  
\Phi(t) \int_0^t \Phi^{-1}(\tau) \NR  
\frac{\partial \rt(\tau)}{\partial \pr_v}  \, 
d\tau \qquad \text{for all } t \ge 0.
\eeq

To provide a more elegant exposition in what follows, we consider the
extension of~\rref{varparv} to the entire response coefficient matrix
$\Rs(\cdot) = \frac{\partial \s}{\partial \pr}(\cdot)$.  Since
$\s_d(t) = \Lio \s_i(t) + \T$ for all $t \geq 0$ and 
$\frac{\partial \T}{\partial \pr_v} =0$, 
\rref{varparv} gives
\beq
\label{dep-dec}
\frac{\partial \s_d}{\partial \pr_v}(t) = \Lio   
\Phi(t) \int_0^t \Phi^{-1}(\tau) \NR \frac{\partial \rt(\tau)}{\partial \pr_v} 
\, d\tau \qquad \text{for all } t \ge 0.
\eeq
Together, \rref{varparv} and~\rref{dep-dec} give
\beq
\label{pre-dec}
\frac{\partial \s}{\partial \pr_v}(t) = \Li  
\Phi(t) \int_0^t \Phi^{-1}(\tau) \NR \, 
\frac{\partial \rt(\tau)}{\partial \pr_v} 
d\tau \qquad \text{for all } t \ge 0.
\eeq
As for the  flux-response coefficients,
\beqn
\frac{\partial \rt}{\partial \pr_v}(t) = \frac{\partial \rt(t)}{\partial \s}
\Li  \Phi(t) \int_0^t \Phi^{-1}(\tau) 
\NR \frac{\partial \rt(\tau)}{\partial \pr_v} 
\, d\tau  + \frac{\partial \rt(t)}{\partial \pr_v} 
\qquad \text{for all } t \ge 0.
\eeqn

\subsection{Scaled Coefficients}

Computations in MCA are typically done using ``scaled'' coefficients: 
derivatives are taken with respect to the logarithms of the variables, giving
relative (rather than absolute) responses.  These scaled sensitivities have 
the advantage that they are dimensionless.  Further, they allow direct 
comparison of responses at different states or across different parameters.

The scaled responses are related
to their unscaled counterparts through multiplication by a ratio of the values
involved.  These relationships can be elegantly described at the matrix level.
Following~\cite{hofmeyr-nutshell}, we define the diagonal matrices
\beqn
\Ds(\cdot) := \text{diag} \s(\cdot) \qquad 
\Dv(\cdot) := \text{diag} \rt(\s(\cdot), \pr, \cdot) \qquad 
\Dp := \text{diag} \pr
\eeqn
whose entries are given by the coefficients of the corresponding vectors.  The
scaled response matrices, denoted with a ``tilde'' ($\tilde{\phantom{a}}$) are 
then given by
\beqn
\bf{\tilde{R}^s}(\cdot) &=& (\Ds(\cdot))^{-1} \Dp \Rs(\cdot) = 
\frac{\partial \ln \s(\cdot)}{\partial \ln \pr} \\
\bf{\tilde{R}^v}(\cdot) &=& (\Dv(\cdot))^{-1} \Dp \Rv(\cdot) =
\frac{\partial \ln \rt(\cdot)}{\partial \ln \pr}.
\eeqn

\section{MCA Theorems}
\label{thms}

We next consider generalizations of the main results of Metabolic
Control Analysis to the case of time-varying sensitivities.  These results 
(the Summation and Connectivity Theorems, and the resulting 
Control-Matrix Equation)
are usually stated in terms of 
{\em control coefficients} and {\em elasticities}, into which the 
response coefficient matrices are factored 
(\cite{reder88, hofmeyr-nutshell}).  
The time varying
responses defined above do not, in general, allow such a factorization.  
However, the responses can be interpreted as the result of applying
an appropriately defined control {\em operator} to an elasticity.  In 
this sense, the standard MCA results can be extended to the time-varying case.
Moreover, we will show that under
certain assumptions on the derivatives of the rates, one can recover 
direct generalizations in terms of matrix factorizations (as was done 
in~\cite{reder-heinrich}).

\subsection{Control Operators}

Given a trajectory of system~\rref{sys}, 
we define the {\em substrate-} and {\em parameter-elasticities}
 of the system as 
\beqn
{\bf \ve}_{\s}(t) := \frac{\partial \rt}{\partial \s}(t) \qquad
{\bf \ve}_{\pr}(t) := \frac{\partial \rt}{\partial \pr}(t) \qquad
\text{for all } t \geq 0.
\eeqn
These vectors describe component (or ``local'') sensitivities of the
isolated reactions associated with the system.

The steady state Summation and Connectivity Theorems admit
 elegant mathematical
descriptions due to the fact that the steady state 
concentration and flux responses can each
be decomposed into a product of two terms:  a matrix of {\em control 
coefficients}
and a vector of parameter elasticities.  It is clear from
equation~\rref{pre-dec} that the general time-varying response cannot
be expressed as a product involving the parameter elasticities (since this
time-varying term appears inside the integral).  However, the relation
between response and elasticity can be expressed as the action
of a control {\em operator} on the elasticity vector, as follows.  (An 
{\em operator}, in this case, is an object that maps functions
to functions.  An introduction to operators appears in the appendix.)

\bd{cco}
The {\em concentration control operator} $\Cs$ maps 
$m$-vector valued functions 
to $n$-vector valued functions as follows.
Given a continuous
$m$-vector valued function $y(\cdot)$ defined on the non-negative reals,
$\Cs(\cdot)(y(\cdot))$ is an $n$-vector valued 
function defined by
\beqn
\Cs(t)(y(\cdot)) := \Li  
\Phi(t) \int_0^t \Phi^{-1}(\tau) \NR \, y(\tau) \, d\tau \qquad 
\text{for all } t \geq 0.
\eeqn
\ed

In a like manner we define the flux control operator.

\bd{fo}
The {\em flux control operator} $\Cv$ maps 
$m$-vector valued functions 
to $m$-vector valued functions as follows.
Given a continuous
$m$-vector valued function $y(\cdot)$ defined on the non-negative reals,
$\Cv(\cdot)(y(\cdot))$ is an $m$-vector valued
function defined by
\beqn
\Cv(t)(y(\cdot)) := {\bf \ve}_{\s}(t)
\Cs(t)(y(\cdot))
+ y(t)
\qquad 
\text{for all } t \geq 0.
\eeqn
\ed
The elasticities and control operators can be scaled in the same manner as the
response coefficients.  Each of the results described below has an analogous
statement in terms of scaled quantities, determined simply by including the
scaling factors.

With these definitions in hand, the generalizations of
the {\em partitioned response properties} (\cite{kas-burns, hofmeyr-nutshell}) follow immediately.  If the parameter vector is such that the
parameters only affect the rates (i.e.~$\pr=\pr_v$), then
\beqn
\Rs(t) = \Cs(t)({\bf \ve}_\pr(\cdot)) \qquad
\Rv(t) = \Cv(t)({\bf \ve}_\pr(\cdot)) \qquad \text{for all } t \geq 0.
\eeqn

We next indicate how the Summation and Connectivity Theorems can be 
interpreted in the light of these definitions.  We will also show how 
the action of these
operators reduces to matrix multiplication when the system is
at steady state.

\subsection{The Summation Theorem}

The Summation Theorem can be stated in terms of the control operators as 
follows.

\bt{Sumrv} {\bf (Summation Theorem)}
If the $m$-vector ${\bf k}$ lies in the nullspace of $\N$ 
(i.e.~$\N {\bf k} = 0$, and so $\NR {\bf k} = 0$ as well), then (using the
simplified notation described in the appendix)
\beqn
\Cs(t)({\bf k}) =0 \qquad \text{and} \qquad \Cv(t)({\bf k}) = 
{\bf k} \qquad \text{for all } t \geq 0.
\eeqn
\ets

\bpr
The result follows from the definitions of the control operators.  
If $\NR {\bf k} = 0$, then
\beqn
\Cs(t)({\bf k}) = \Li  
\Phi(t) \int_0^t \Phi^{-1}(\tau) \NR \, {\bf k} \, d\tau =0 \qquad 
\text{for all } t \geq 0,
\eeqn
and
\beqn
\Cv(t)({\bf k}) = \ve_\s \Cs(t)({\bf k}) + {\bf k} = {\bf k}
\qquad 
\text{for all } t \geq 0.
\eeqn
\epr

To state the result in a more standard form, we allow the control
operators to act on matrices as follows.  If ${\bf r}_1(\cdot), 
\ldots, {\bf r}_n(\cdot)$ are $m$-vector valued functions, then we 
interpret
\beqn
\Cs(t)([{\bf r}_1(\cdot) \ldots {\bf r}_n(\cdot)]) := \left [
\Cs(t)({\bf r}_1(\cdot)) \ldots \Cs(t)({\bf r}_n(\cdot)) \right ].
\eeqn
With this notation, a  
direct corollary of the Summation Theorem is the following.
\bc{sumc}
If the matrix ${\bf K}$ has columns which form a basis for the nullspace of
$\N$, then
\beqn
\Cs(t)({\bf K}) = 0 \qquad \text{and} \qquad \Cv(t)({\bf K}) = 
{\bf K} \qquad \text{for all } t \geq 0.
\eeqn
\ecs
\qed

The Summation Theorem is normally stated as a property of the
{\em control coefficients} of a system at steady state.  However, the steady
state version can be given
an equivalent formulation which does not make reference to 
control coefficients.  

\bp{Sumr} {(\bf Summation Theorem -- steady state)}
Suppose the system is at steady state, so we may drop time dependencies 
on all terms.  If the parameter vector  is 
such that the parameters only affect the rates (i.e. $\pr = \pr_v$) and
 each column of the matrix
${\bf \ve_p}=\frac{\partial \rt}{\partial \pr}$ 
lies in the nullspace of $\N$, then 
\beqn
\Rs =0 \qquad \text{and} \qquad \Rv = {\bf \ve_p}.
\eeqn
\eps\qed

Stated in this way, the Theorem can be extended directly to 
time-varying responses.  From Theorem~\ref{Sumrv}, we have the following.

\bc{Sumrvo} {\bf (Summation Theorem -- alternative statement)}
If the parameter vector  is such that the parameters only affect
the rates (i.e.~$\pr = \pr_v$) and
 each column of the matrix
${\bf \ve_p}(t)=\frac{\partial \rt}{\partial \pr}(t)$
 lies in the nullspace of $\NR$ for each $t \geq 0$, then 
\beqn
\Rs(t) =0 \qquad \text{and} \qquad \Rv(t) = {\bf \ve_p}(t) 
\qquad \text{for all } t \geq 0.
\eeqn
\ecs\qed

\subsubsection{Special Case: ${\bf \ve_p}$ constant}

In special cases, Theorem \ref{Sumrv} reduces to more familiar variants.
A simplification which can be made to the general case described above is
to assume that some derivatives of the rate law $\rt(\cdot,\cdot, \cdot)$ 
are constant
throughout the evolution of the system.  This was the case considered 
\preprint{by Heinrich and Reder}
 in~\cite{reder-heinrich}, where it
was
argued that for systems near a stable equilibrium,
it may be reasonable to approximate the derivatives of $\rt$ by their values
at the equilibrium.  

If one restricts to the special case where the parameter elasticity
${\bf \ve_p}$ is constant 
throughout the evolution of the system,  the control operators need not
be defined to act on arbitrary functions, but only on constant $m$-vectors.  
The action of the operators can
in this case be described simply as matrix multiplication, e.g. for any
constant $m$-vector ${\bf y}$,  
\beqn
\Cs(t)({\bf y}) = \left [ \Li  
\Phi(t) \int_0^t \Phi^{-1}(\tau) \NR  \, d\tau 
\, \right ] {\bf y} \qquad 
\text{for all } t \geq 0.
\eeqn

In this case, through a slight abuse of notation, 
time varying {\em control coefficients} can be defined by
\beqn
\Cs(t)& := &\Li \Phi(t) \int_0^t \Phi^{-1}(\tau) \NR \, d\tau \\
\Cv(t)& := &\ve_{\s}(t) \Li \Phi(t) \int_0^t \Phi^{-1}(\tau) \NR \, d\tau
 + {\bf I}_m \\
\eeqn

Here, we can state a more standard Summation Theorem:  if
the matrix ${\bf K}$ has columns which form a basis for the null space of
$\N$ (equivalently of $\NR$), then for all $t \geq 0$,
\beqn
\Cs(t) {\bf K} &=& 0 \\
\Cv(t) {\bf K} &=& {\bf K},
\eeqn
where the left-hand-side is interpreted as a matrix product.

\comment{  example of summation theorem

\subsubsection{An Illustration}

Consider the simple branched pathway indicated in Figure~\ref{sumfig}.
Suppose the the rates are given in terms of three external parameters 
$k_1$, $k_2$, $k_3$ by the equations
\beqn
v_1 := k_1 \qquad v_2 := k_2 S \qquad v_3 := k_3.
\eeqn
This can be interpreted as a system for which reaction 3 is saturated.  
To ensure a physically meaningful steady state, we assume
$k_1 > k_3$.  The stoichiometry matrix for the system is 
\beqn
\N = \left [ \begin{array}{ccc} 1 & -1 & -1 \end{array} \right ].
\eeqn
Take nominal values of the parameters as $k_1^0$, $k_2^0$, and $k_3^0$
We denote by $p$ the parameter $\frac{k_1}{k_1^0} + \frac{k_3}{k_3^0}$.
Then 
\beqn
\frac{\partial \rt}{\partial p} = \left [ \begin{array}{c} 1 \\ 0 \\ 1 
\end{array} \right ],
\eeqn
which is constant, and is in the nullspace of $\N$.  Thus, {\em
regardless of the initial condition and resulting trajectory of $S$}, the
responses of the concentration of $S$ and the rates to a perturbation
in $p$ will be as described in the statement of the Summation Theorem {\em
for all time}.
} 

\subsection{Connectivity Theorem}

The Connectivity Theorem for control operators is as follows.

\bt{ct}{\bf (Connectivity Theorem)}  Applying the control operators to
the function 
${\bf \ve}_\s(\cdot) \Li = \frac{\partial \rt}{\partial \s}(\cdot) \Li$ yields,
for all $t \ge 0$,
\beqn
\Cs(t)({\bf \ve}_\s(\cdot) \Li) &=& \Li (\Phi(t) - {\bf I}_{n_0})  \\
\Cv(t)({\bf \ve}_\s(\cdot) \Li) &=& {\bf \ve}_\s(t) \Li 
\Phi(t).
\eeqn
\ets
While the right-hand-sides of the equations above may not look 
familiar, the reader should note that if $\Phi(t)=0$ then these reduce to more
standard connectivity relations.  In what follows we will consider a case where
$\Phi(t)$ approaches zero as steady state is reached, so that the classical MCA
result is recovered.

\bpr
The definition of the concentration control operator gives
\beq
\Cs(t)(\frac{\partial \rt}{\partial \s}(\cdot) \Li) = \Li  
\Phi(t) \int_0^t \Phi^{-1}(\tau) \NR \, 
\frac{\partial \rt(\tau)}{\partial \s} \Li \, d\tau 
\qquad \text{for all} t \ge 0. \label{coneq}
\eeq
From equation~\rref{homo}, we have that
\beq
\label{phid}
\left [\frac{d}{dt} \Phi(t) \right ]\Phi^{-1}(t) = 
\NR \frac{\partial \rt(t)}{\partial \s}
\Li \qquad \text{for all } t \geq 0.
\eeq
Moreover, from the fact that
\beqn
0= \frac{d}{dt} {\bf I}_{n_0} = \frac{d}{dt} [ \Phi(t) \Phi^{-1}(t) ] =
\left [ \frac{d}{dt} \Phi(t) \right ] \Phi^{-1}(t) + \Phi(t) \left 
[ \frac{d}{dt} \Phi^{-1}(t) \right ],
\eeqn
we see that
\beqn
\frac{d}{dt} \Phi^{-1}(t) = - \Phi^{-1}(t) \left [ \frac{d}{dt} \Phi(t)
\right ] \Phi^{-1}(t) \qquad \text{for all } t \geq 0.
\eeqn
Thus, from~\rref{coneq} and~\rref{phid},
\beqn
\Cs(t)(\frac{\partial \rt}{\partial \s}(\cdot) \Li) &=&\Li  
\Phi(t) \int_0^t \Phi^{-1}(\tau) \NR \, 
\frac{\partial \rt(\tau)}{\partial \s} \Li \, d\tau \\
&=& \Li  
\Phi(t) \int_0^t \Phi^{-1}(\tau) \left [ \frac{d}{d\tau} \Phi(\tau) \right ] 
\Phi^{-1}(\tau) \, d\tau \\
&=&
- \Li  
\Phi(t) \int_0^t \frac{d}{d\tau} \Phi^{-1}(\tau) \, d\tau
\qquad \text{for all } t \ge 0.
\eeqn
Application of the Fundamental Theorem of Calculus gives
\beqn
\Cs(t)(\frac{\partial \rt}{\partial \s}(\cdot) \Li)
 &=& - \Li \Phi(t) [\Phi^{-1}(t) 
- \Phi^{-1}(0) ]\\
&=& \Li ( \Phi(t) - {\bf I}_{n_0}) \qquad \text{for all } t \ge 0,
\eeqn
as $\Phi^{-1}(0) = {\bf I}_{n_0} = \Phi(0)$.
With the definition of the flux control operator, we conclude
\beqn
\Cv(t)(\ve_{\s}(\cdot) \Li) &=& 
\ve_{\s}(t) \Li ( \Phi(t) - {\bf I}_{n_0})
+ \ve_{\s}(t) \Li \\
&=& \ve_{\s}(t) \Li \Phi(t)
\qquad \text{for all } t \ge 0.
\eeqn
\epr

\comment{general matrix statement of connectivity
\bc
Given any matrix ${\bf M}$, 
\beqn
\Cs(t)({\bf \ve}_s(t) \Li {\bf M}) = 
\Li (\Phi(t) - {\bf I}_m) {\bf M} \\
\Cv(t)(\frac{\partial \rt}{\partial \s}(\cdot) \Li {\bf M}) = 
\frac{\partial \rt}{\partial \s}(t) \Li 
\Phi(t) {\bf M}.
\eeqn
In particular, if an $m$-vector valued function ${\bf r}(t)$ lies in the column space
of the matrix valued function ${\bf \ve}_s(t) \Li$, then ${\bf r}(t) = {\bf \ve}_s \Li {\bf M}$ for some vector ${\bf M}$, and so the action of the operators
on ${\bf r}(\cdot)$ is given as above.
\ec

\bpr
This follows immediately from the fact that the action of the linear control
operators on matrices is defined in terms of its actions on column vectors.
\epr
} 

The Connectivity Theorem is typically stated as a property of the steady
state 
control coefficients.  As is the case for the Summation Theorem, 
an equivalent formulation can be given which does not make reference to 
control coefficients, as follows.  

\bp{Conr} {\bf (Connectivity Theorem -- steady state)} Suppose 
the system is at steady state, so we may drop time dependencies 
on all terms.  If the parameter vector  is 
such that the parameters
only affect the rates (i.e.~$\pr = \pr_v$) and the elasticities satisfy 
$\ve_{\pr} =
{\bf \ve}_{\bf s} \Li$, then
\beqn
\Rs = - \Li \qquad \text{and} \qquad \Rv = 0.
\eeqn
\eps\qed

There is a direct extension to the general case.  From Theorem~\ref{ct}, we 
have the following.

\bc{Conrv} {\bf (Connectivity Theorem -- alternative statement)}
If the parameter vector  is 
such that the parameters
only affect the rates (i.e.~$\pr = \pr_v$) and the elasticities satisfy 
$\ve_{\pr}(t) = 
{\bf \ve}_{\bf s}(t) \Li$ 
for each $t \geq 0$, then
\beqn
\Rs(t) = \Li (\Phi(t) - {\bf I}_{n_0} ) \qquad \text{and} 
\qquad \Rv(t) = {\bf \ve}_{\bf s}(t) \Li \Phi(t)
\qquad \text{for all } t \geq 0.
\eeqn
\ecs\qed

\subsubsection{Special Case: $\ve_{\s}$ and
$\ve_{\pr}$ constant}

Again, in the special case where the derivatives of $\rt(\cdot, \cdot, \cdot)$
are constant throughout the trajectory, the analysis simplifies.

If $\ve_{\s}= \frac{\partial \rt}{\partial \s}$ is constant 
throughout the evolution of the system, the fundamental matrix 
defined by equation~\rref{homo} can be described explicitly as the matrix 
exponential
\beqn
\Phi(t) = e^{\NR \ve_{\s} \Li t}
 \qquad \text{for all } t \geq 0.
\eeqn
In this case~\rref{varpar} provides an explicit formula for the 
sensitivity coefficients.

Heinrich and Reder considered the case in which
both $\ve_{\pr}$ and 
$\ve_{\s}$ are constant along the trajectory of the
system in~\cite{reder-heinrich}.  
In that case, the control operators again need only be defined
on constant vectors, their action reducing to matrix multiplication.
That is, given any constant $m$-vector ${\bf y}$, 
\beq
\nonumber
\Cs(t)({\bf y}) &=& \left [ \Li  
e^{\NR \ve_{\s} \Li t}
 \int_0^t  e^{-\NR \ve_{\s} \Li \tau}
\NR \, d\tau \right ] {\bf y} \\
&=& \left [ \Li  
 \int_0^t  e^{\NR \ve_{\s} \Li (t-\tau)} \NR
\, d\tau \right ] {\bf y} \qquad \text{for all } t \ge 0.
\label{exp-int}
\eeq
The matrix $\NR \ve_{\s} \Li$ is the Jacobian of 
equation~\rref{redsys}.  In the case where it is nonsingular, \rref{exp-int}
further reduces to
\beqn
\Cs(t)({\bf y})
&=& \left [
\Li (e^{\NR \ve_{\s} \Li t} - {\bf I}_{n_0}) 
(\NR \ve_{\s} \Li)^{-1}
\NR \right ]{\bf y} \qquad \text{for all } t \ge 0.
\eeqn
This assumption of nonsingularity is rather standard 
(see, e.g.~\cite{reder88}); in particular, it holds  
when the analysis is being carried out near an asymptotically 
stable equilibrium point.

Under these assumptions the control coefficients take the form of the 
time-varying matrices
\beqn
\Cs(t)& := & \Li (e^{\NR \ve_{\s} \Li t} - 
{\bf I}_{n_0})
(\NR \ve_{\s} \Li)^{-1}  \NR \\
\Cv(t)& := & \ve_{\s} \Li (e^{\NR \ve_{\s} \Li t} - 
{\bf I}_{n_0})
(\NR \ve_{\s} \Li)^{-1}  \NR + {\bf I}_m,
\eeqn
and the Connectivity Theorem reduces to the form presented 
in~\cite{reder-heinrich}:  for all $t \ge 0$,
\beqn
\Cs(t) {\bf \ve}_\s  \Li &= &
\Li (e^{\NR {\bf \ve}_\s \Li t} - {\bf I}_{n_0}) \\
\Cv(t) {\bf \ve}_\s \Li &=& 
{\bf \ve}_\s \Li 
e^{\NR {\bf \ve}_\s \Li t}.
\eeqn

If such a a trajectory further satisfies the condition that it is approaching 
an asymptotically stable steady state, then it will be the case that
\beqn
\lim_{t \into \infty} e^{\NR {\bf \ve}_\s \Li t} =0.
\eeqn
Then, taking limits, we find that the control coefficients reduce to their
standard forms~\cite{reder88} as the steady state is approached,
\beqn
\lim_{t \into \infty} \Cs(t) &=& 
-\Li (\NR {\bf \ve}_\s \Li)^{-1}  \NR \\
\lim_{t \into \infty} \Cv(t) &= &
-{\bf \ve}_\s 
\Li (\NR {\bf \ve}_\s \Li)^{-1}  \NR + {\bf I}_m.
\eeqn
The time-varying Summation and Connectivity 
Theorems described above reduce to the standard statements in this case, as
described in~\cite{reder-heinrich}.

\subsection{Control Matrix Equation}

Together, the steady state 
Summation and Connectivity Theorems provide a relationship
between the elasticities and control coefficients.  An 
elegant description of this relation is the Control Matrix
Equation~\cite{hofmeyr-nutshell}.  This equation has a direct
generalization in terms of the control operators described above.

\bc{con-mat-eq}
If ${\bf K}$ is a matrix whose columns form a basis for the nullspace of
$\N$, then, making use of the notation for vectors of operators introduced
in the appendix, 
\beqn
\left [ \begin{array}{c} \Cv(t) \\ \Cs(t) \end{array} \right ]
\left ( \begin{array}{cc} {\bf K} & - {\bf \ve}_\s \Li \end{array} \right )
=\left [ \begin{array}{cc} {\bf K} & - {\bf \ve}_\s(t) \Li \Phi(t) \\ 
0  & \Li ({\bf I}_{n_0} - \Phi(t)) \end{array} \right ]
\qquad \text{for all } t \ge 0.
\eeqn
\ec

\section{Applications and a Caveat for Oscillating Systems}

We present applications of the above analysis to three models.  The first 
is a ``toy'' model -- a  stable two-species pathway.
  This analysis provides a 
straightforward illustration of the interpretation of time-varying 
response coefficients.  
The second example is a simple oscillating system. In light of this example we
provide a brief discussion of the interpretation of sensitivity  
coefficients of oscillatory systems.  Finally, we provide a
``realistic'' example -- an analysis of a phototransduction model based on
work of 
Reike and Baylor~\cite{baylor}.  
Here the power of a time-varying sensitivity analysis is demonstrated, 
since we are
able to make statements about the effect of parameter variations on
the critical transient behaviour displayed by the model.

\subsection{Basic Pathway}
Consider the simple pathway shown in Figure~\ref{path}.
\preprint{\begin{figure}
\begin{center}
\epsfysize=55pt
\centerline{\epsffile{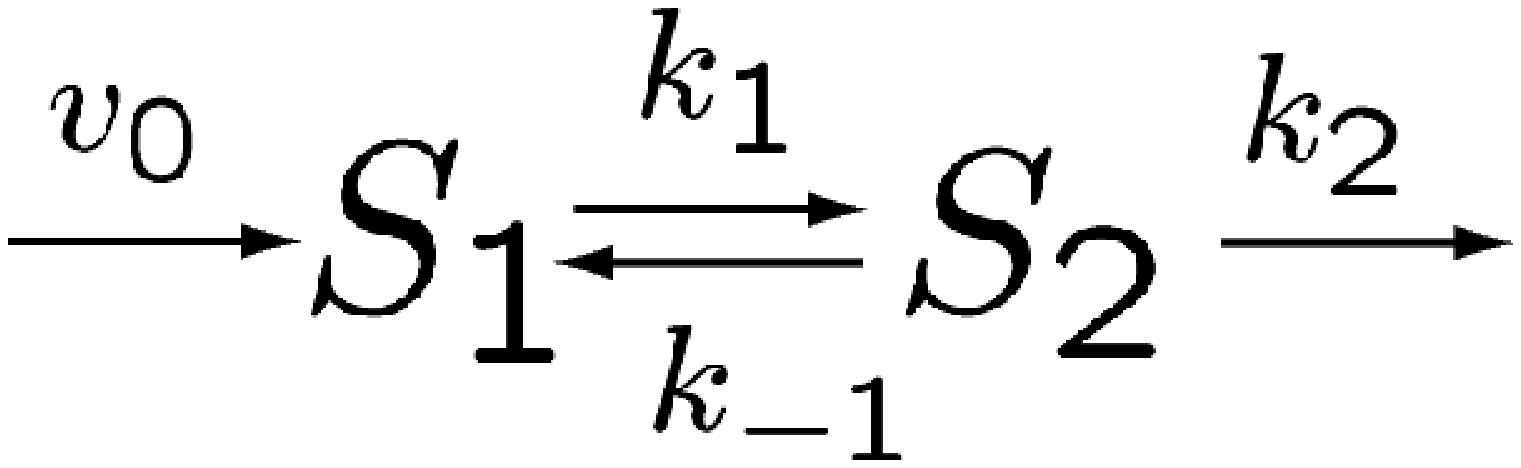}}
\end{center}
\caption{Pathway}
\label{path}
\end{figure}}
Assume mass-action kinetics for each of the reactions.
We will model the system with the following parameter values: $v_0 = 4$, 
$k_1 = 3$, $k_{-1} = 0.5$, and $k_2 = 2$.  Since the flow into $S_1$ is
fixed, the system is asymptotically stable to the equilibrium $(S_1, S_2) =
(\frac{5}{3}, 2)$.  We will consider the unscaled response coefficients defined
for initial conditions at the steady state (recall $\Rs(\cdot)$ depends on 
the nominal parameters values and the initial state).  Thus in this case
the nominal trajectory will be the steady state trajectory
$(S_1(t), S_2(t)) = (\frac{5}{3}, 2)$ for all $t \geq 0$.

\comment{
First consider the effect of a perturbation in $v_0$, as shown in 
Figure~\ref{simple-v0}.
Interpretation is immediate, the effect of the perturbation takes time
to propagate through the system; the response of $S_2$ is delayed
in comparison to that of $S_1$, since it takes longer for a change in $v_0$
to ``reach'' $S_2$; and eventually, the sensitivities reach the steady state
values. 
\preprint{\begin{figure}
\epsfysize=175pt
\centerline{\epsffile{matlab/simple-v0.eps}} 
\caption{Perturbation in $v_0$}
\label{simple-v0}
\end{figure}}
} 

First consider the sensitivity to a perturbation in $k_1$ (shown in 
Figure~\ref{simple-k1}).
\preprint{\begin{figure}
\epsfysize=175pt
\centerline{\epsffile{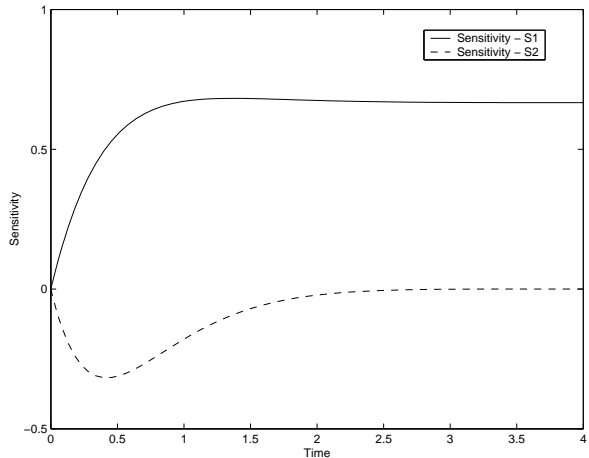}} 
\caption{Perturbation in $k_1$}
\label{simple-k1}
\end{figure}}
The steady state response of $S_2$ to such a perturbation is nil (since at
steady state, $S_2 = \frac{v_0}{k_2}$).  However, the analysis shows 
(and intuition concurs), that an increase in $k_1$ will produce a 
transient increase in $S_2$, which fades gradually as  $S_1$ decreases to  
accommodates the perturbed parameter.  The response in $S_1$ grows to 
reach its steady state value -- a small overshoot can be seen at about time 
$t=1$.   Not surprisingly, the situation for
perturbations in $k_{-1}$ is similar (but reversed), 
with a transient {\em decrease} in 
$S_2$ and a positive steady state response in $S_1$ (Figure~\ref{simple-kn1}).
\preprint{\begin{figure}
\epsfysize=175pt
\centerline{\epsffile{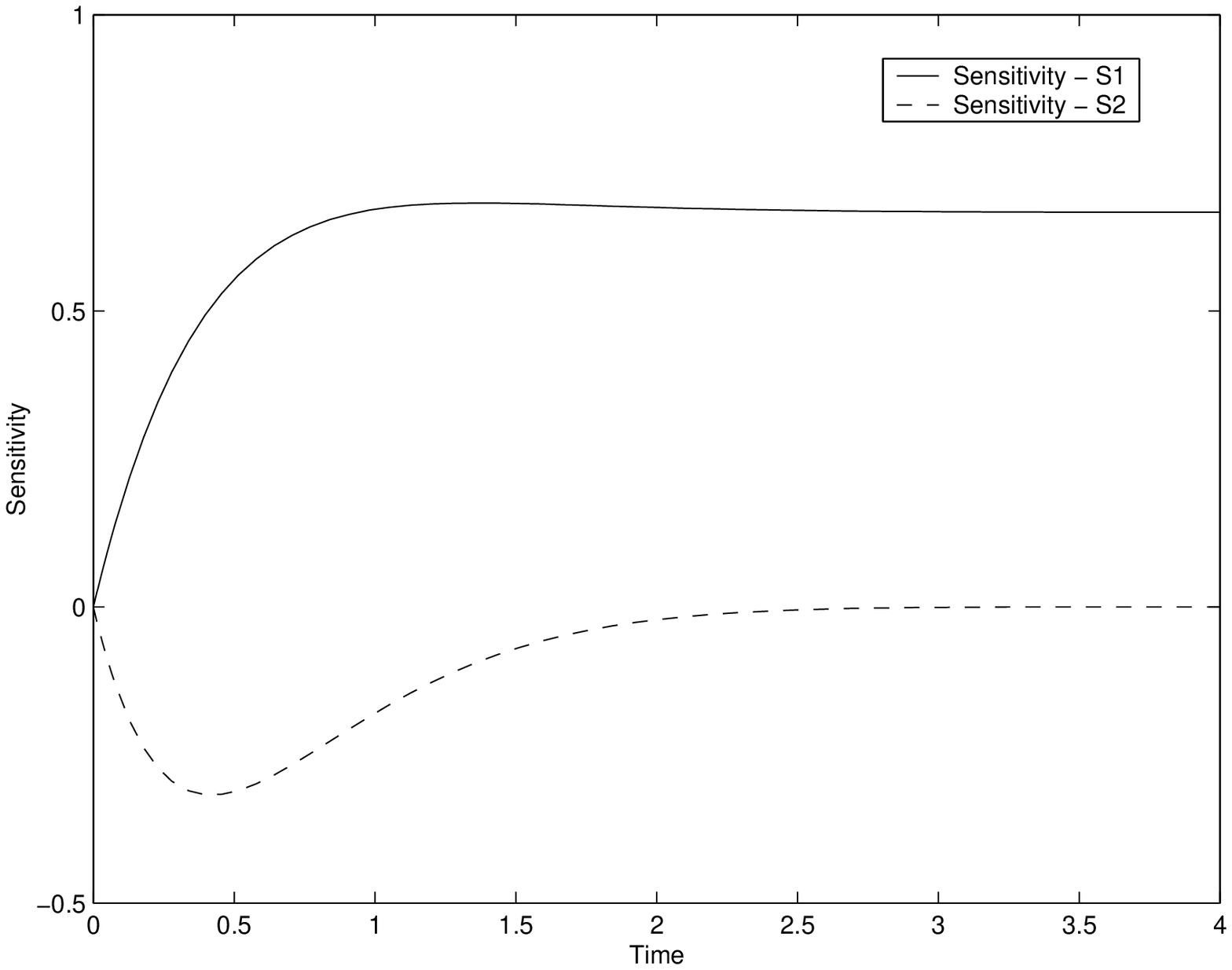}} 
\caption{Perturbation in $k_{-1}$}
\label{simple-kn1}
\end{figure}}
One can read from the graphs how the effect on each concentration changes over 
time.  For instance, the effect of these changes on $S_2$ is felt most
strongly at about $t=0.4$.

Finally, we consider the effect of a perturbation in the initial value of
$S_1$ (Figure~\ref{simple-s1}).
\preprint{\begin{figure}
\epsfysize=175pt
\centerline{\epsffile{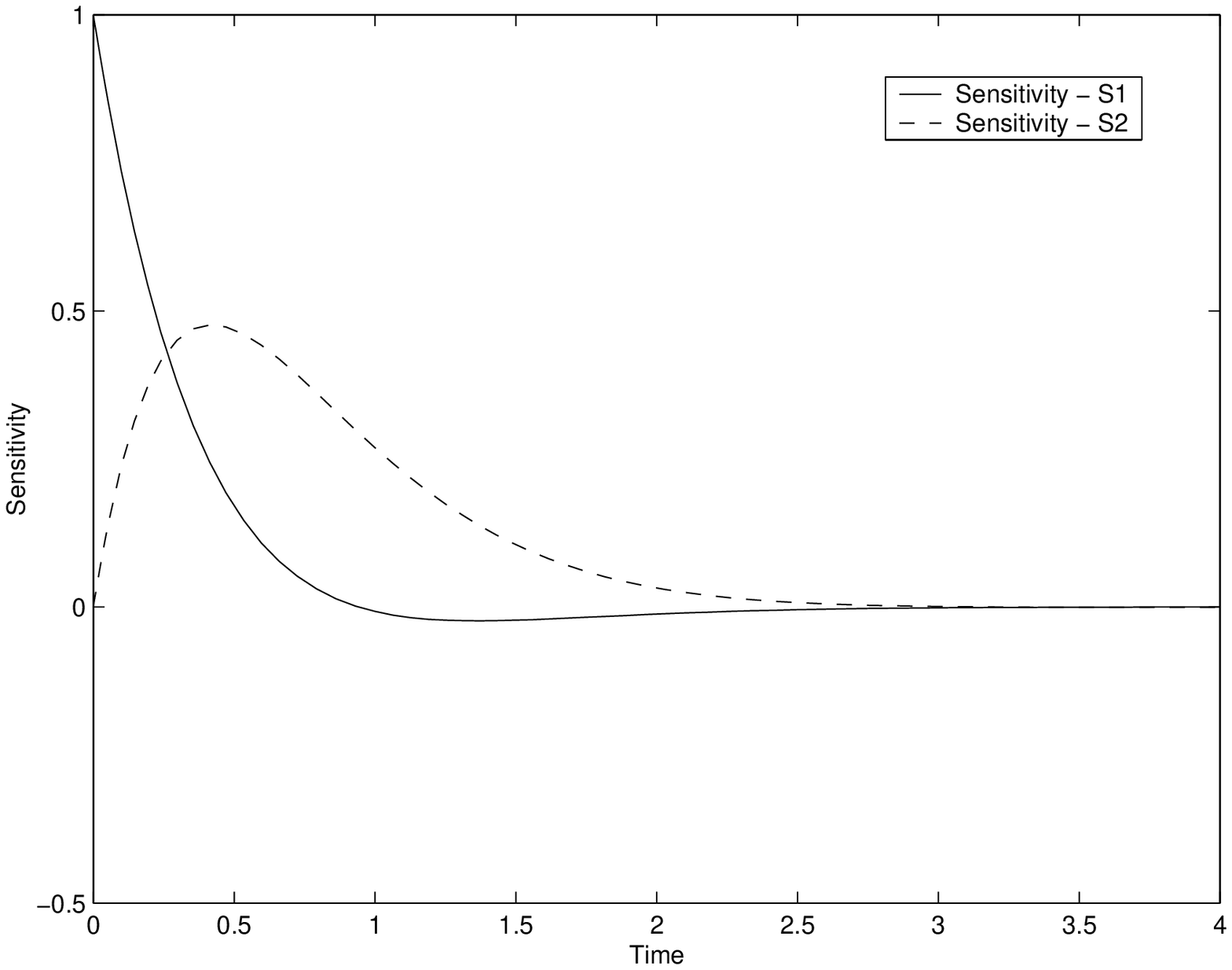}} 
\caption{Perturbation in $S_1(0)$}
\label{simple-s1}
\end{figure}}
Again, the results are as expected, with transient effects of the 
perturbations ``washing through'' the pathway -- the result being no effect on
the steady state solution.

Having considered the effect of perturbations around a steady state,
we now turn our attention to a system whose time-behaviour is more complex.

\subsection{Oscillatory Systems}

We consider another simple two species pathway, based on a model
which has been employed in investigations of glycolytic 
oscillations~\cite{hrr77}.  
The network is shown
in Figure~\ref{osc}.  
\preprint{\begin{figure}
\epsfysize=100pt
\centerline{\epsffile{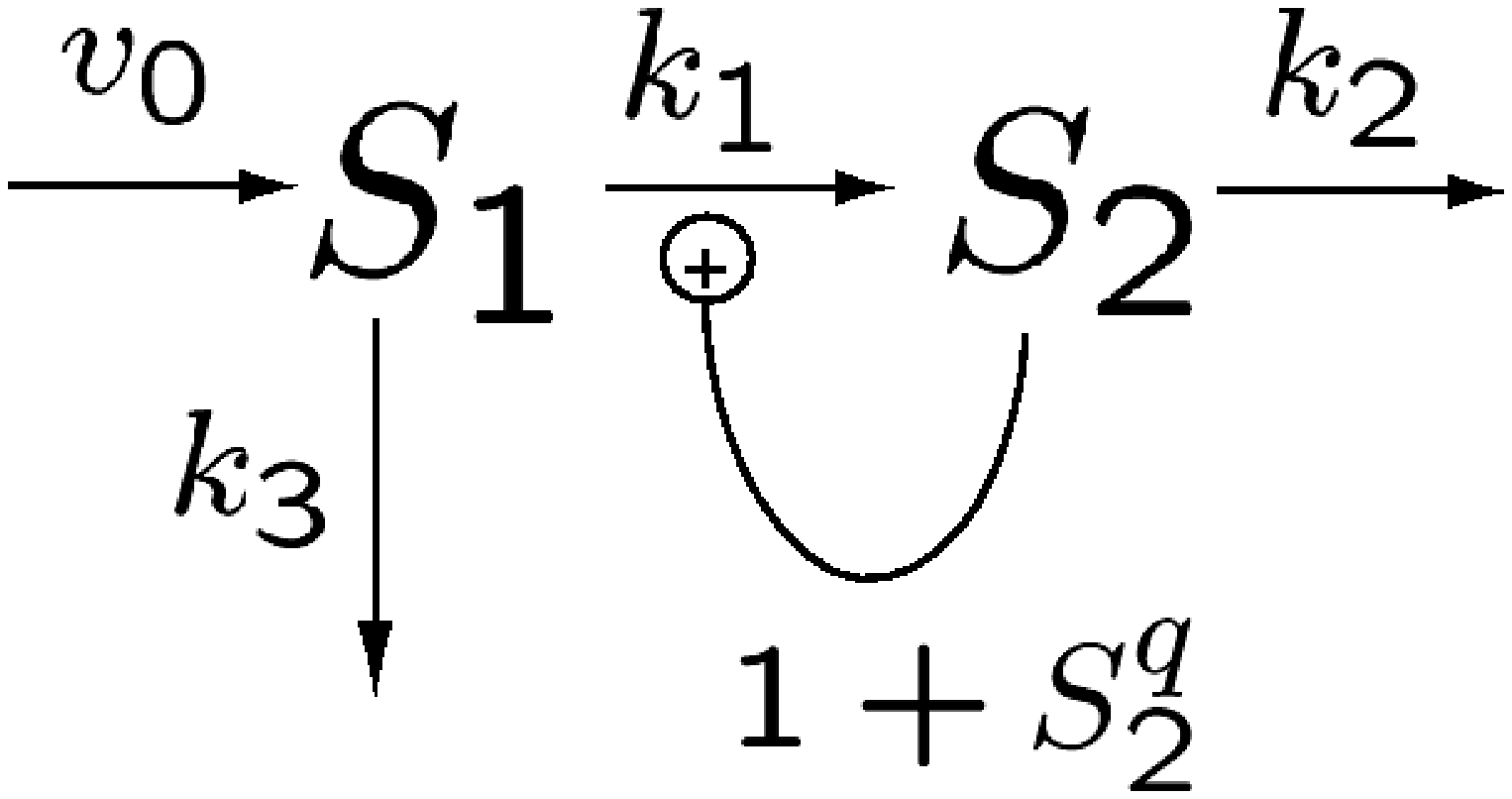}} 
\vspace{1cm}
\caption{Oscillatory System}
\label{osc}
\end{figure}}
The rates are calculated by mass-action, with the activation by $S_2$ appearing
as a multiplicative factor (i.e. the rate of production of $S_2$ is given by 
$k_1 S_1 (1 + S_2^q)$).
For parameters in a certain range, the 
concentrations of $S_1$ and $S_2$ follow oscillatory trajectories; the 
positive feedback from $S_2$ drives the oscillation.  

We choose nominal parameter values of $v_0 = 8$, $k_1 = 1$, $k_2=5$, $k_3=1$,
and $q=3$, and nominal initial values of $(S_1(0), S_2(0)) = 
(1,3)$.  The oscillatory behaviour is a limit cycle (cf.~\cite{hrr77}), 
and so any initial
conditions in a neighbourhood of these points will yield similar
trajectories.  

We first consider perturbations in initial values.  The limit
cycle behaviour is stable; after a 
perturbation of the initial conditions, the system will tend to the limit
cycle as time tends to infinity.  However, unlike the previous case where convergence to the same behaviour meant  
that the response shrank to zero as time moved on, for this model
(and for systems with limit cycles in general), the {\em phase} of the 
oscillations depends on the initial values.  Thus
the perturbed trajectory is out of phase with the nominal trajectory.  The 
result is a periodic response -- oscillating with the same period as the limit 
cycle and indicating the difference
in the trajectories at each point in time.  
The response for perturbations in
$S_1(0)$ is shown in Figure~\ref{osc-s1}, 
along with the nominal trajectory.
\preprint{\begin{figure}
\epsfysize=175pt
\centerline{\epsffile{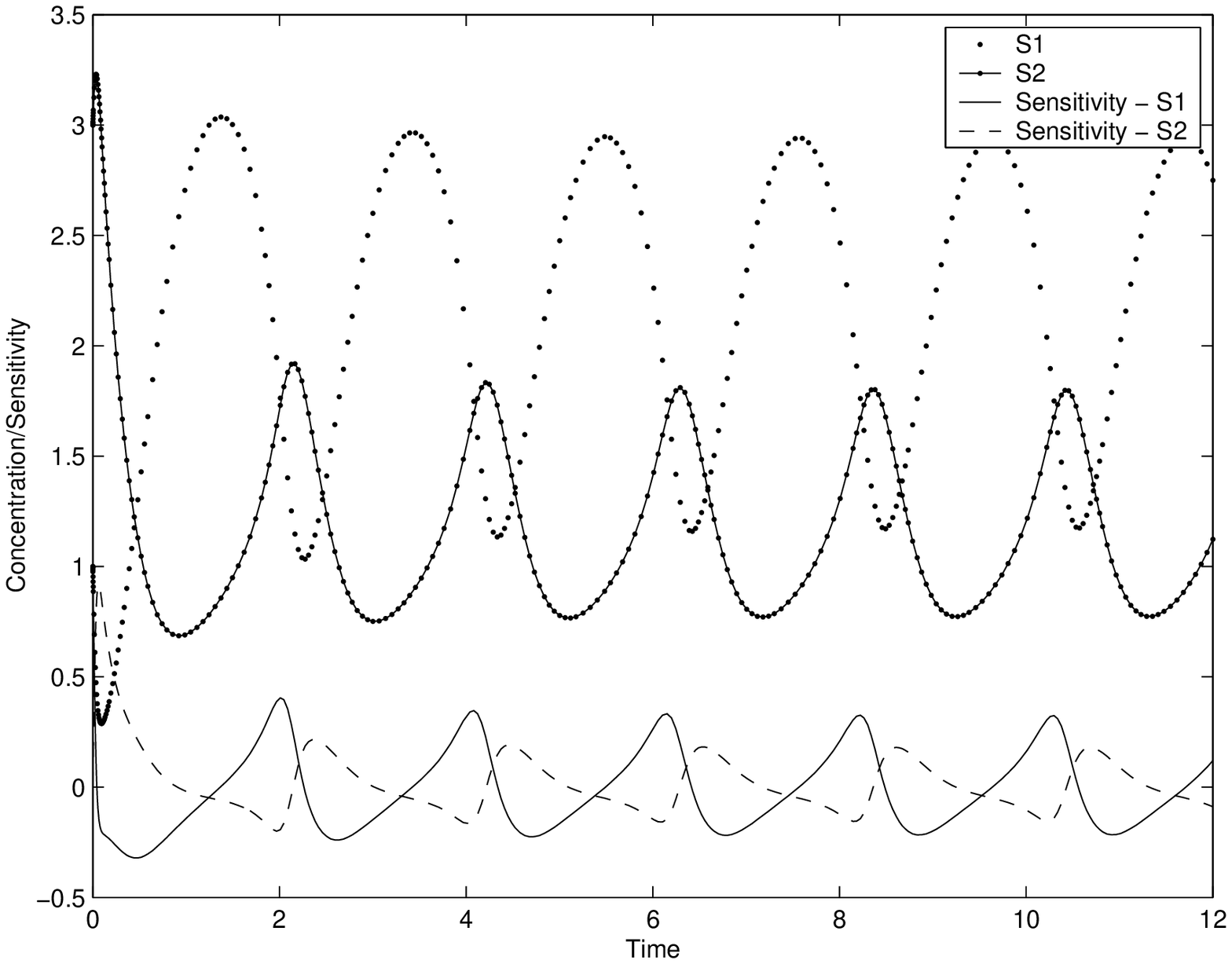}} 
\caption{Perturbation in $S_1(0)$}
\label{osc-s1}
\end{figure}}
\comment{
\begin{figure}
\epsfysize=175pt
\centerline{\epsffile{matlab/osc-s2.eps}} 
\caption{Perturbation in $S_2(0)$}
\label{osc-s2}
\end{figure}
} 
The response for perturbations in $S_2(0)$ (not shown) is similar.

Care must be taken in extending this set of ideas to perturbations of
{\em kinetic parameters} of an oscillatory system.  In general, the results
of such analysis may not be useful.  The response coefficients for 
perturbations in $v_0$ are shown (again, along with the nominal trajectories)
in Figure~\ref{osc-v0}.
\preprint{\begin{figure}
\epsfysize=175pt
\centerline{\epsffile{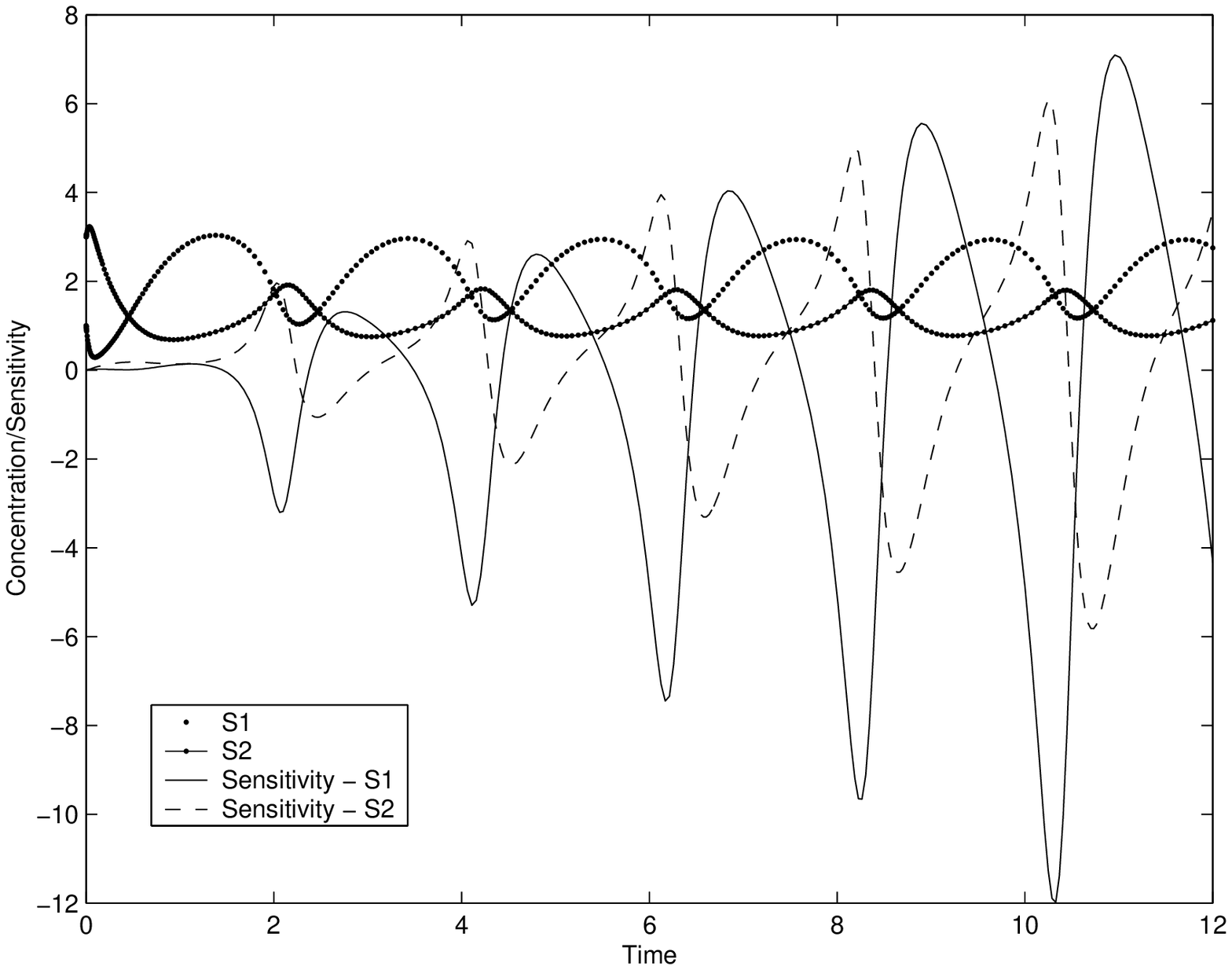}} 
\caption{Perturbation in $v_0$}
\label{osc-v0}
\end{figure}}
While the sensitivities match the period of the system in a quasi-periodic
behaviour, they
oscillate more and more widely, and continue to do so as time tends to
infinity.  As a result, these responses
are not especially useful when trying to predict the effect of
finite perturbations on a real system.  The trouble, as pointed out 
in~\cite{demin, Kho-periodic}, is that after a perturbation in
a kinetic parameter, the system again exhibits a limit cycle behaviour,
but this time with a {\em different period}.  When two trajectories follow
a similar cycle with different periods, they will always reach a point when
they are completely out of phase,  no matter how small the difference
in their periods.  As the perturbations get smaller, the difference in
period shrinks as well, so that it takes longer and longer for this ``point of 
maximum difference'' to be reached.  However, as long as the perturbation is 
nonzero, that point will always be achieved.  Since the response coefficient
is defined as the limit of the ratio of the difference in trajectories to 
the size of the perturbation as the perturbation goes to zero, we find that 
as time tends to infinity, the response approaches this
``maximum difference'' divided by zero, that is it diverges.

In~\cite{demin} and~\cite{Kho-periodic}, some clever strategies have been 
devised for interpreting the sensitivities of oscillatory systems.  
Kholodenko {\em et al.} treat the case of 
asymptotically stable systems under the influence of periodic 
external forces in~\cite{Kho-periodic}.  
Such systems exhibit limit cycles whose period is set
by the external force.  In this case, 
perturbations in kinetic parameters do not
result in changes in the period, and so the response coefficients are more
meaningful. 
In~\cite{demin}, \preprint{Demin {\em et al.} treat} perturbations of 
autonomously oscillating systems \journal{are treated}
by measuring the response in the {\em Fourier
coefficients} of the trajectory.  Fourier control coefficients are
introduced, which 
give a useful interpretation of sensitivities for any periodic systems.

Having considered the interpretation of response coefficients for steady state
and oscillating systems, we now treat the most interesting case -- 
a system in which the
behaviour of interest occurs during transients.

\subsection{Phototransduction}

Phototransduction is the process by which organisms convert 
light signals into nerve signals.  As in all signal transduction
pathways, the steady state behaviour of the system is less interesting 
than its transient response.  We will illustrate
a sensitivity analysis of the single photon response 
in a simple model of phototransduction loosely based on previous modeling work 
in~\cite{baylor} and~\cite{pugh}.  

The response 
mechanism is modelled as follows.  The system is at steady state in the
absence of light input.  A photon of
light activates a rhodopsin molecule 
to start the system response.  Activated rhodopsin 
(Ra) in 
turn activates a heterotrimeric G protein (G).  
The activated G protein $\alpha$-subunit (Ga) 
dissociates from its beta-gamma subunits ($\beta \gamma$) and then
binds with a cGMP phosphodiesterase (PDE) to form the active PDEa-Ga complex.  
The PDE enzyme catalyzes the degradation of cGMP in the cell.  Activation by 
Ga greatly enhances the catalytic activity of PDE, resulting in a reduction
in cellular levels of cGMP.  The cGMP-gated ion channels in the cell membrane
then close, causing a net efflux of calcium ion
(Ca), a net outward current (J), and
a corresponding hyperpolarization of the cell's membrane. 

The deactivation phase of the response
is modelled as follows.  Activated G protein $\alpha$ subunit, either free 
(Ga) or bound (PDEa-Ga), is converted to the deactivated form (Gd).
Finally, the deactivated G protein (Gd) 
recombines with the beta-gamma subunit ($\beta \gamma$) to form the
heterotrimeric G protein (G). 

The model incorporates four independent variables (G, Ga, PDEa-Ga, and cGMP)
along with five dependent variables (Gd, PDE, $\beta \gamma$, Ca, and J) 
and several
parameters as indicated below.   Note, J denotes 
not the actual current but rather the magnitude of the deviation
from the resting current (DarkJ).
The input to the system is the time-varying
level of Ra. 
The network is shown in Figure~\ref{photo}.
\preprint{\begin{figure}
\epsfysize=160pt
\centerline{\epsffile{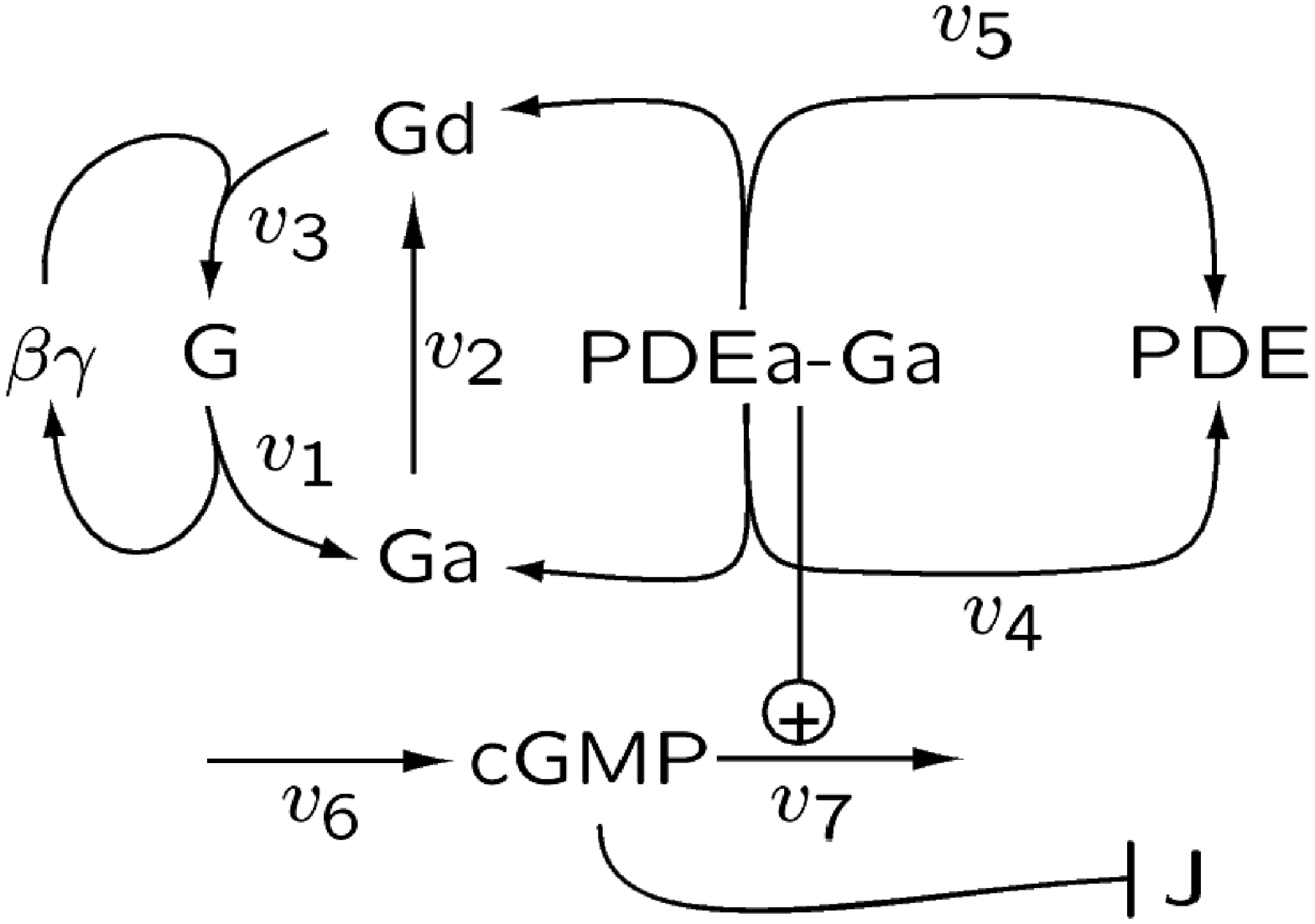}} 
\vspace{0.5cm}
\caption{Phototransduction Pathway}
\label{photo}
\end{figure}}
The dynamics are given by
\beqn
v_1 & = &k_{Ga} \text{Ra} \\
v_2 & = & k_{Gd} \text{Ga} \\
v_3 & = & k_{G1} \text{$\beta \gamma$ Gd} \\
v_4 & = & k_{PDE1} \text{Ga PDE} - k_{PDE1m} \text{PDEa-Ga} \\
v_5 & = & k_{PDE2} \text{PDEa-Ga} \\
v_6 & = & \frac{k_{cG}}{1 + \left (\frac{\text{Ca}}{K_g} \right )^q} \\
v_7 & = & k_{cGd0} \text{cGMP} + k_{cGd} \text{cGMP PDEa-Ga} \\
\\
\text{Gd} &=& \text{Gtot} - \text{G} - \text{Ga} \\
\text{PDE} &=& \text{PDEtot} - \text{PDEa-Ga} \\
\text{$\beta \gamma$}& =& \text{$\beta \gamma$tot} - \text{G} \\
\text{Ca}& =& \frac{k_{C_i}}{k_{C_o}} 
\left (\frac{\text{cGMP}}{K_{0.5}} \right )^3 \\
\text{J} & = & \text{DarkJ} 
\left( 1-  \left (\frac{\text{cGMP}}{K_{0.5}} \right )^3 \right )
\eeqn

For simulations, we choose nominal parameter values and initial conditions
as follows
\beqn
\begin{array}{ll}
k_{Ga} = 1.0 \text{ms}\!^{-1} & k_{Gd} = 0.005 \text{ms}\!^{-1}\\
k_{G1} = 0.002 \text{ms}\!^{-1} 
\text{molecules}\!^{-1} \, \mu\!\!\text{m}\!\!^2 &
k_{PDE1} = 0.0003 \text{ms}\!^{-1}  
\text{molecules}\!^{-1} \, \mu\!\!\text{m}\!\!^2 \\
k_{PDE1m} = 0.000001 \text{ms}\!^{-1}&
k_{PDE2} = 0.005 \text{ms}\!^{-1}  \\
k_{cG} = 0.058 \text{mM} \text{ms}\!^{-1} &
K_g = 0.2 \text{mM} \\
q= 2.0 & 
k_{cGd0} = 0.002 \text{ms}\!^{-1} \\
  k_{cGd} = 0.000004 \text{ms}\!^{-1} 
\text{molecules}\!^{-1} \, \mu\!\!\text{m}\!\!^2 &
k_{C_o} = 1.44  \\ k_{C_i} = 0.72 & K_{0.5} = 4 \text{mM}\\
\\
\text{Gtot} = 3000 \text{molecules} \mu\!\!\text{m}\!\!^{-2}
& \text{PDEtot} = 500 \text{molecules} \mu\!\!\text{m}\!\!^{-2} \\
\text{$\beta \gamma$tot} = 3000 \text{molecules} \mu\!\!\text{m}\!\!^{-2}

& \text{DarkJ} = 12.0 \text{pA} \\
\\
\text{G}\!(0) = 3000 \text{molecules} \mu\!\!\text{m}\!\!^{-2} & 
\text{Ga}\!(0) = 0 \text{molecules} \mu\!\!\text{m}\!\!^{-2} \\
\text{PDEa-Ga}\!(0) = 0 \text{molecules} \mu\!\!\text{m}\!\!^{-2} & 
\text{cGMP}\!(0) = 4.0 \text{mM}
\end{array}
\eeqn

The single activated rhodopsin molecule excited by a single photon is
represented by a square 100ms pulse,
\beqn
\text{Ra}(t) = \left \{ \begin{array}{ll} 0 & 0 \le t \le 100 \\
                                          1 & 100 < t \le 200 \\
                                          0 & 200 < t \end{array} \right. ,
\eeqn
measured in$\text{molecules} \mu\!\!\text{m}\!\!^{-2}$.

We first consider the effect of a perturbation in the parameter $k_{Ga}$, which
describes the effect of activated rhodopsin on the activation of G.  
The nominal trajectories for G, Ga and PDEa-Ga 
are shown in Figure~\ref{nomG} and Figure~\ref{nomGa}. 
\preprint{\begin{figure}
\epsfysize=175pt
\centerline{\epsffile{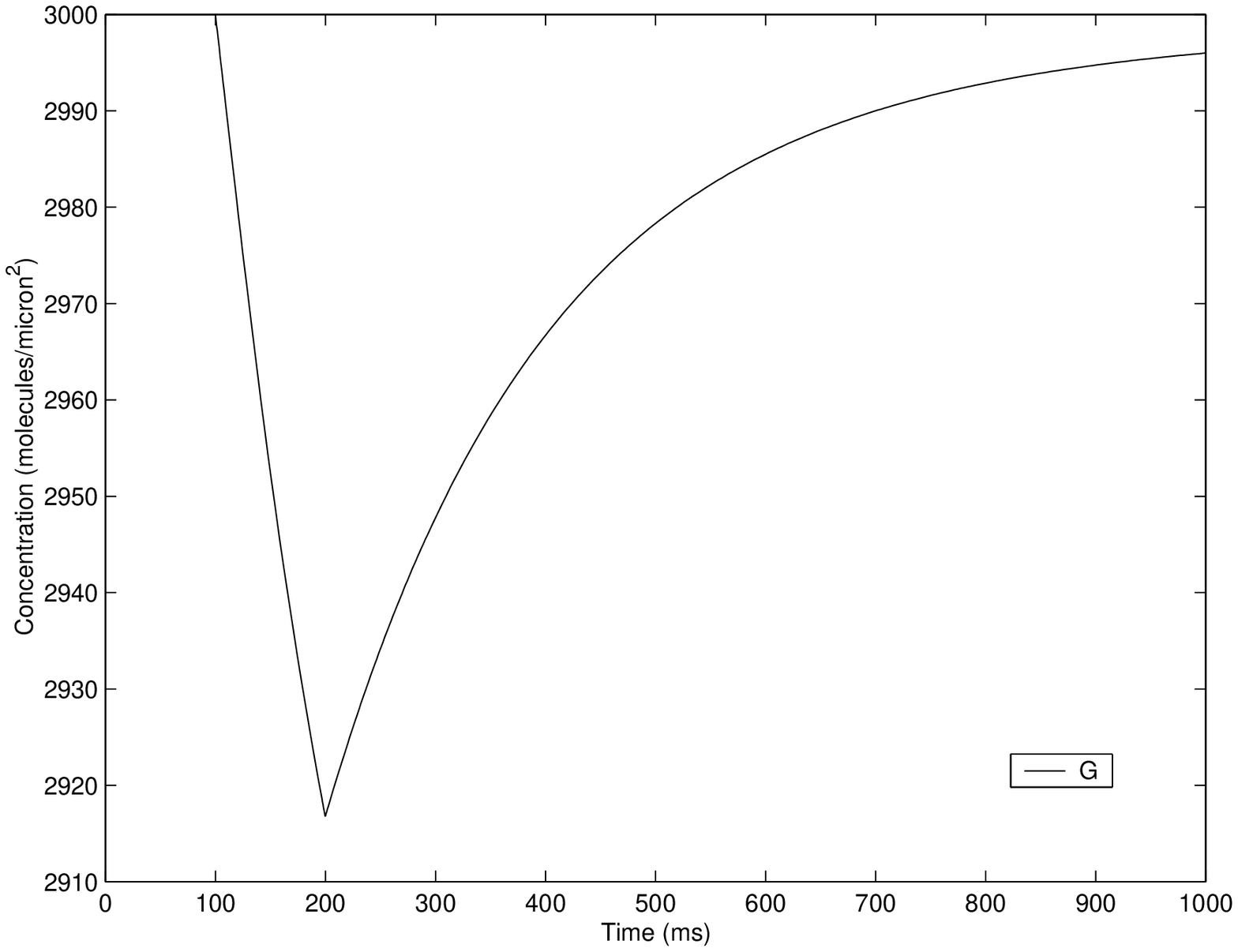}} 
\caption{Nominal trajectory for G}
\label{nomG}
\end{figure}
\begin{figure}
\epsfysize=175pt
\centerline{\epsffile{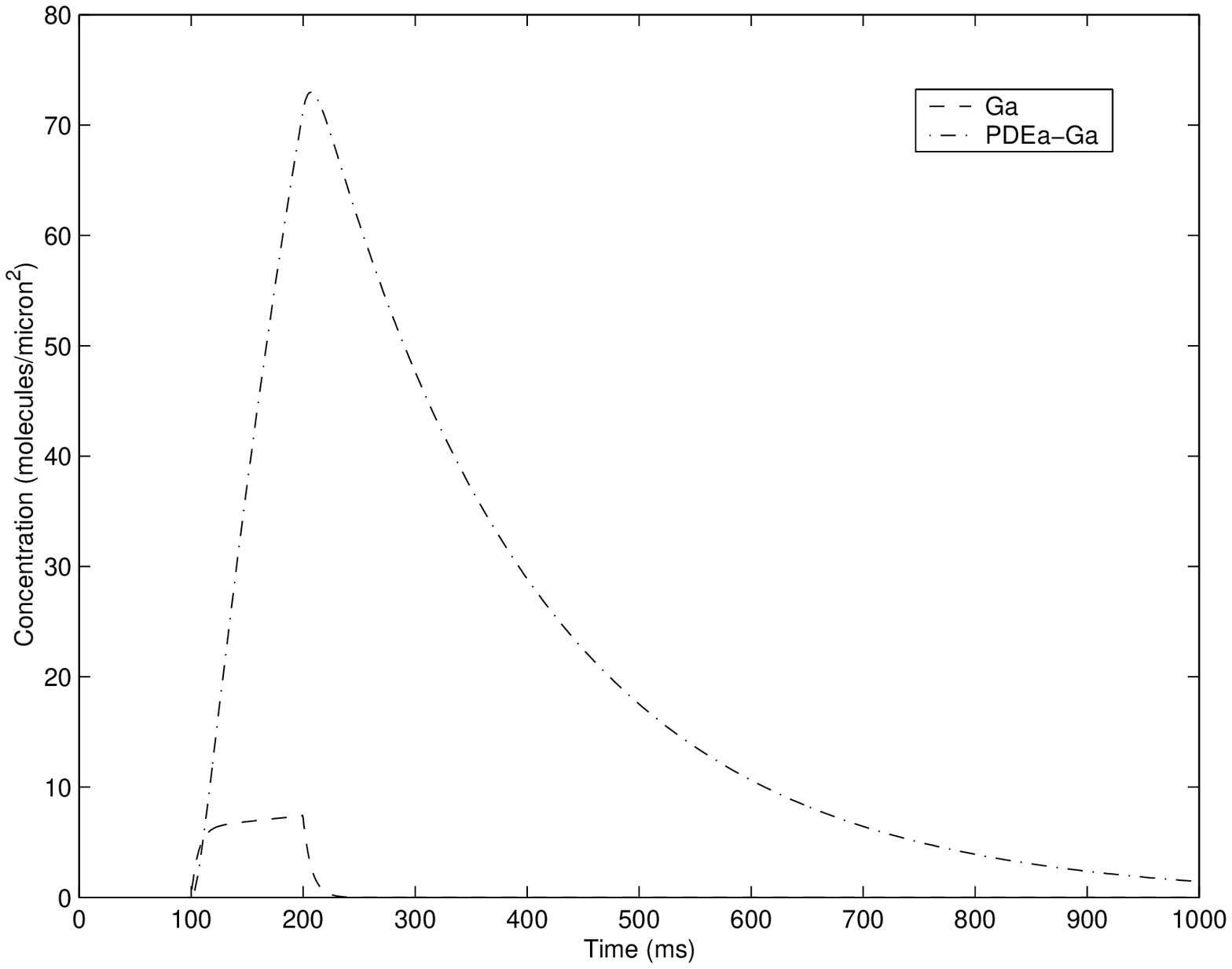}} 
\caption{Nominal trajectory for Ga, PDEa-Ga}
\label{nomGa}
\end{figure}}
(There is no
activity until time $t=100$ since the system is at steady state while the
input is zero).  Figure~\ref{senk} 
\preprint{\begin{figure}
\epsfysize=175pt
\centerline{\epsffile{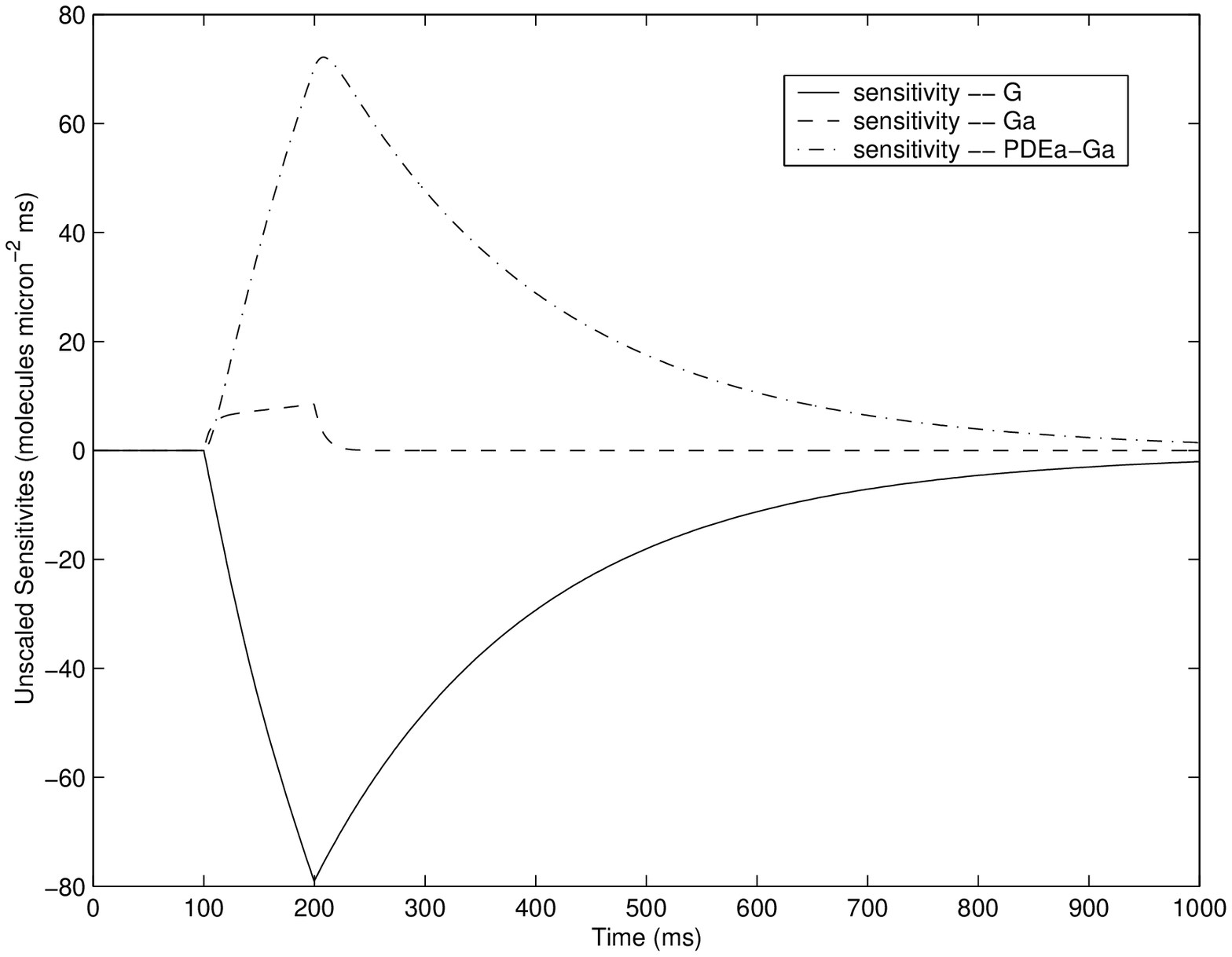}} 
\caption{Absolute sensitivities to perturbations in $k_{Ga}$}
\label{senk}
\end{figure}}
shows the absolute (unscaled) 
sensitivity of G, Ga and 
PDEa-Ga to perturbations in $k_{Ga}$.  As expected, an increase in $k_{Ga}$ 
causes 
a decrease in the level of G and an increase in the levels of Ga and 
PDEa-Ga.  These 
responses track the trajectories themselves, with the largest response being
observed when each species is farthest from its ``resting'' (steady) state.

The primary signal of interest in this system is the level of current (J)
produced by the input.  The time history of the current produced by 
the nominal input and parameter values is shown in 
Figure~\ref{nomJ}.
\preprint{\begin{figure}
\epsfysize=175pt
\centerline{\epsffile{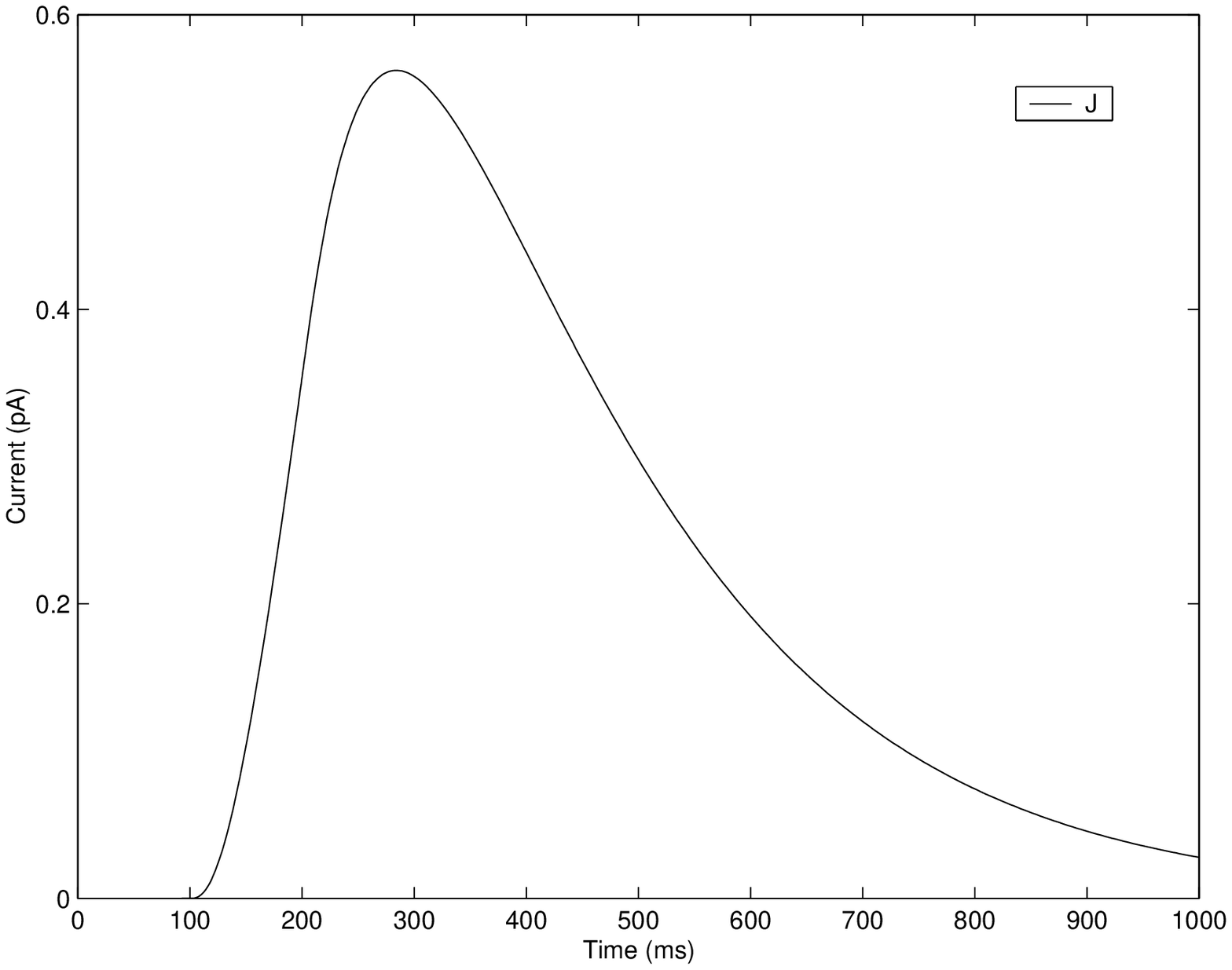}} 
\caption{Nominal trajectory of J}
\label{nomJ}
\end{figure}}
The effect of perturbations in three different parameters on J are shown
in Figure~\ref{senJcom}.  
Since this analysis is meant to compare the effects of
the changes, the relative (i.e.~scaled) responses are shown.
\preprint{\begin{figure}
\epsfysize=175pt
\centerline{\epsffile{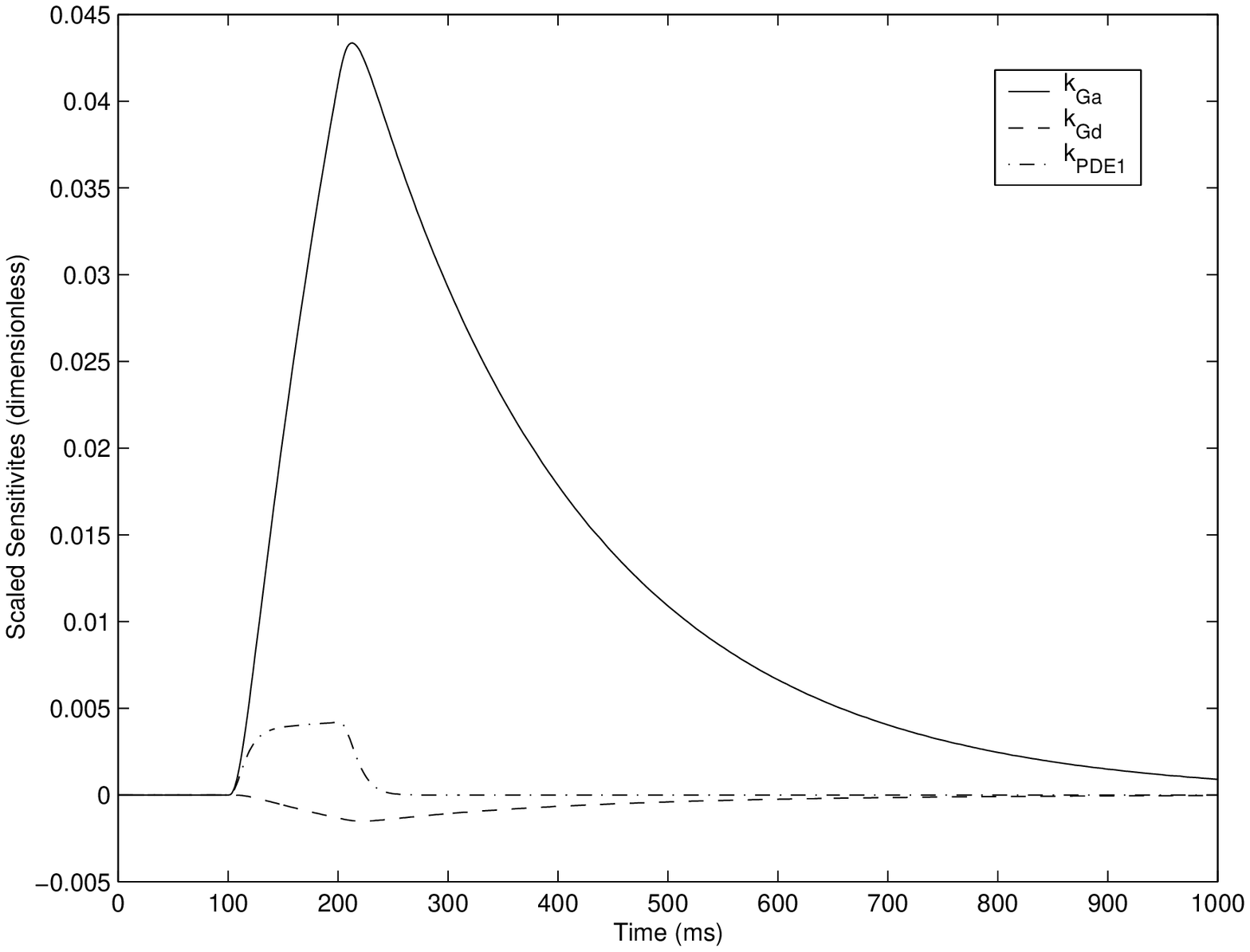}} 
\caption{Relative sensitivities of J}
\label{senJcom}
\end{figure}}
The relative strengths of perturbations in the three parameters $k_{Ga}$,
$k_{Gd}$ and $k_{PDE1}$ are immediate.  For instance, to increase the
response in J, it is clear that an increase in $k_{Ga}$ will
have far more impact than an equal (relative) increase in $k_{PDE1}$ or
decrease in $k_{Gd}$.


\section{Conclusion}

Since its introduction, Metabolic Control Analysis has proven to be a 
useful tool in discovering the 
distribution of control in biochemical systems at steady state.  
In this paper, the computation and interpretation
of system sensitivities has been considered for arbitrary trajectories.
The
 main results of MCA (Summation and Connectivity Theorems) have been shown
to have a valid interpretation not just at steady state, but 
throughout the system dynamics.  In
analyzing the distribution of control for systems in which transients
are of interest, the time varying sensitivity may prove to be a 
valuable tool
in determining system behaviour.

\noindent{{\bf Acknowledgment}} The authors would like to thank Tau-Mu Yi for
providing the phototransduction model and for many useful discussions.

\appendix
\section{Appendix: A Primer on Operators}

A {\em function} from $\R$ into $\R$ is defined as a rule which assigns 
to each real number\footnote{The symbol $\R$ denotes the set of
real numbers.} a unique real number in its range, for example, the
function $h(\cdot)$ defined by $h(t)=t^2$ for each $t \in \R$.  This can be
generalized by allowing functions to act on pairs of numbers, or $n$-tuples
in general (e.g. $g(x,y,z) = x^2 - yz$).  

The concept of a function can be extended to yield
an {\em operator}, which takes a function  as its argument.  
For example,
we could consider the operator $T(\cdot)$ defined by integration: 
\beqn
T(f(\cdot)) = \int_0^1 f(s) \, ds
\eeqn
for {\em each function $f(\cdot)$} defined on $\R$ (provided the integral
of $f$ exists).
Define the functions $h$ and $r$ by $h(t)=t^2$ and 
$r(t) = \cos(t)$. Then 
\beqn
T(h(\cdot)) = \int_0^1 s^2 \, ds = \frac{1}{3}, \qquad \text{and} \qquad
T(r(\cdot)) = \int_0^1 \cos(s) \, ds = \sin(1).
\eeqn
Another example of an operator is function evaluation. Define $G$ by  
$G(f(\cdot))= f(5)$ for each
function $f(\cdot)$.  Then $G(h(\cdot)) = 25$ and $G(r(\cdot))= \cos(5)$.

The operators $T$ and $G$ defined above are commonly referred to as 
{\em functionals}, since they map functions to numbers.  More generally,
by allowing an operator to have a second, scalar argument, we can construct
operators which map {\em functions to functions}.  For example, extending
$T$ to depend on a scalar argument $t$, we could define $\hat{T}$ by
\beqn
\hat{T}(t, f(\cdot)) = \int_0^t f(s) \, ds.
\eeqn
An alternative notation is $\hat{T}(t)(f(\cdot)) = \hat{T}(t, f(\cdot))$.
Thus, given a real-valued function, $T$ returns a function defined for
$t \geq 0$:
\beqn
T(t)(h(\cdot)) = \int_0^t s^2 \, ds = \frac{t^3}{3} \qquad \text{and} \qquad
T(t)(r(\cdot)) = \int_0^t \cos(s) \, ds = \sin(t).
\eeqn
As another example, we could modify $G$ by defining 
$\hat{G}(t)(f(\cdot)) = f(2t)$.  Then $\hat{G}(t)(h(\cdot)) = (2t)^2$ and
$\hat{G}(t)(r(\cdot)) = \cos(2t)$, for all $t \in \R$.

As described in the main text, we can extend the action of operators to
row vectors in a component-wise fashion, for example
\beqn
\hat{T}(t)([\begin{array}{cc} h(\cdot) &  r(\cdot) \end{array}]) = 
[\begin{array}{cc} \frac{t^3}{3} &  \sin(t) \end{array}].
\eeqn
Moreover, we can combine operators column-wise, yielding the
notation
\beqn
\left [ \begin{array}{c} \hat{T}(t) \\ \hat{G}(t) \end{array} \right ] 
 ([\begin{array}{cc} h(\cdot) &  r(\cdot) \end{array}]) =
\left [ \begin{array}{cc} \frac{t^3}{3} &  \sin(t) \\
                      (2t)^2 & \cos(2t) \end{array} \right ].
\eeqn

Particular cases of operator actions which appear in the main text are:
\bi
\item  constant functions as arguments to operators.  For example, 
defining the function $k(\cdot)$ by $k(t) = 3$ for all $t \in \R$, we have
\beqn
T(k(\cdot)) = 3, \ \  \hat{T}(t)(k(\cdot)) = 3t.
\eeqn
Using a simplified notation, we can write
\beqn
T(3) = 3, \ \  \hat{T}(t)(3) = 3t;
\eeqn
\item  operators which act by multiplication.  For example, define the
operator $H$ by $H(t)(f(\cdot)) = 3 t f(t)$.  Then
\beqn
H(t)(h(\cdot))  = 3 t^3 \ \ \ \ \ H(t)(r(\cdot)) = 3 t \cos(t) \ \ \ \ \
H(t)(3)  = 9t.
\eeqn
\ei

\preprint{

} 

\journal{

} 

\end{document}